# Ice-free geomorphometry of Queen Maud Land, East Antarctica: 3. Belgica and Yamato (Queen Fabiola) Mountains

I.V. Florinsky[1*], S.O. Zharnova[2]

[1] Institute of Mathematical Problems of Biology, Keldysh Institute of Applied Mathematics, Russian Academy of Sciences, Pushchino, Moscow Region, 142290, Russia

[2] National Research Tomsk State University, 36 Lenin Ave., Tomsk, 634050, Russia

## Abstract

Geomorphometric modeling and mapping of ice-free Antarctic areas can be applied for obtaining new quantitative knowledge about the topography of these unique landscapes and for the further use of morphometric information in Antarctic research. Within the framework of a project of creating a physical geographical thematic scientific reference geomorphometric atlas of ice-free areas of Antarctica, we performed geomorphometric modeling and mapping of two, partly ice-free mountainous areas of the eastern Queen Maud Land, East Antarctica. These include the Belgica Mountains and Yamato (Queen Fabiola) Mountains. As input data, we used two fragments of the Reference Elevation Model of Antarctica (REMA). For the two ice-free areas and adjacent glaciers, we derived models and maps of eleven, most scientifically important morphometric variables (i.e., slope, aspect, horizontal curvature, vertical curvature, minimal curvature, maximal curvature, catchment area, topographic wetness index, stream power index, total insolation, and wind exposition index). The obtained models and maps describe the ice-free topography of the Belgica Mountains and Yamato (Queen Fabiola) Mountains in a rigorous, quantitative, and reproducible manner. New morphometric data can be useful for further geological, geomorphological, glaciological, ecological, and hydrological studies of these areas.

**Keywords:** topography, digital elevation model, mathematical modeling, Antarctica

## 1 Introduction

There are three main types of ice-free areas in Antarctica: (1) Antarctic oases, i.e. coastal, shelf, and mountainous ice-free areas; (2) ice-free islands (or areas thereof) situated outside the ice shelves; and (3) ice-free mountain chains (or their portions) and nunataks (Markov et al., 1970; Simonov, 1971; Korotkevich, 1972; Alexandrov, 1985; Pickard, 1986; Beyer and Bölter, 2002; Sokratova, 2010).

Geomorphometry deals with the mathematical modeling and analysis of topography as well as relationships between topography and other components of geosystems (Evans, 1972; Moore et al., 1991; Wilson and Gallant, 2000; Shary et al., 2002; Hengl and Reuter, 2009; Minár et al., 2016; Florinsky, 2017, 2025a). Geomorphometric modeling and mapping of ice-free Antarctic areas (Florinsky, 2023a, 2023b; Florinsky and Zharnova, 2025a, 2025b, 2025c, 2025d, 2025e) can be used for obtaining new quantitative knowledge about the topography of these terrains and for the further use of morphometric information in solving problems of geomorphology, geology, glaciology, soil science, ecology, and other sciences. A project has been launched to create a physical geographical thematic scientific reference geomorphometric atlas of ice-free areas of Antarctica (Florinsky, 2024, 2025b).

Currently, as part of this project, we carry out a geomorphometric modeling and mapping of ice-free areas of Queen Maud Land, East Antarctica. Previously, we published series of geomorphometric maps for the Sôya Coast and the Prince Olav Coast of that land (Florinsky and Zharnova, 2025b, 2025c). In this paper, we present results for two, partly ice-free mountainous areas of the eastern Queen Maud Land, namely, the Belgica Mountains and the Yamato (Queen Fabiola) Mountains.

* Correspondence to: iflor@mail.ru

## 2 Study areas

The Belgica and Yamato (Queen Fabiola) Mountains are isolated, relatively small, partly ice-free mountain massifs located in the eastern Queen Maud Land, East Antarctica (Fig. 1, Table 1). The mountains are surrounded by glaciers on all sides.

The Belgica Mountains are situated inland about 240 km southeast of the Princess Ragnhild Coast. The Yamato (Queen Fabiola) Mountains are located inland about 170 km southwest of the Prince Harald Coast and about 200 km southeast of the Princess Ragnhild Coast. From the west, the Belgica Mountains are bounded by the Sør Rondane Glacier. The Yamato (Queen Fabiola) Mountains lie northeast of the Belgica Mountains at a distance of about 190 km; the two mountain systems are separated by the Belgica Glacier. From the east, the Yamato (Queen Fabiola) Mountains are bounded by the glacial Mizuho Plateau. The vast icefields of Thorshavnheiane, a large inland region of Dronning Maud Land, lie south of the Belgica and Yamato (Queen Fabiola) Mountains. In particular, the blue-ice Inseki Icefield is located southward the Yamato (Queen Fabiola) Mountains (Fig. 1).

The Belgica Mountains were discovered by the Belgian Antarctic Expedition in 1958. The Belgian team reached the mountains by air from the Base Roi Baudouin, conducted a topographic survey, and named the mountains after RV *Belgica* of the Belgian Antarctic Expedition of 1897–1899. In 1979, a traverse party of the 20th Japanese Antarctic Research Expedition (JARE) visited the region and performed topographic and geological surveys (Kojima et al., 1981, 1982).

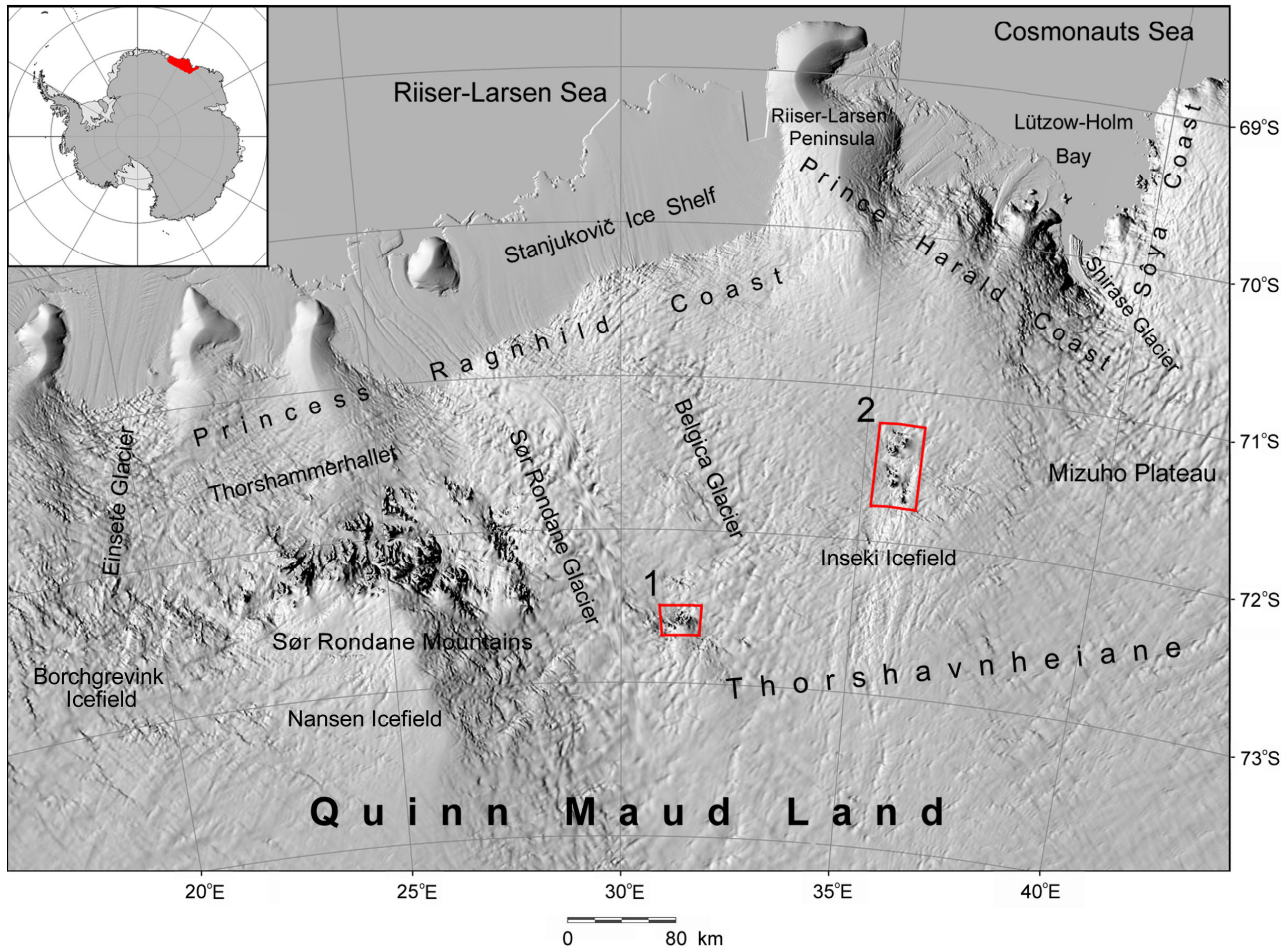

**Fig. 1** Geographical location of the study areas in the eastern Queen Maud Land: 1—Belgica Mountains, 2—Yamato (Queen Fabiola) Mountains. The hill-shaded map was produced by the REMA Explorer (PGC, 2022–2024). Geographical names were selected from three sources (Geographical Survey Institute, 1998; Korotkevich et al., 2005; Norwegian Polar Institute, 2017).

**Table 1** Characteristics of the study areas (Fig. 1) and their DEMs.

| # | Area name | Geographical coordinates* | Ice-free area**, km$^2$ | DEM size, m | DEM size, point | Points with elevation values |
|---|---|---|---|---|---|---|
| 1 | Belgica Mountains | 72.58374° S, 31.24304° E | 32.9 | 34,440 × 35,480 | 1,722 × 1,774 | 3,047,833 |
| 2 | Yamato (Queen Fabiola) Mountains | 71.51580° S, 35.67792° E | 116.5 | 36,351 × 63,680 | 1,136 × 1,990 | 2,260,640 |

* The geographic coordinates of the conditional centers are given in the WGS84 system.
** Ice-free areas were estimated from satellite images (Fig. 2).

The Yamato (Queen Fabiola) Mountains were discovered from air by the Norwegian Antarctic Expedition in 1937. In October 1960, the Belgian Antarctic Expedition obtained oblique air photographs of the mountains and named them after Queen Fabiola of Belgium. In November–December 1960, the region was first visited by a traverse party of the 4th JARE. Its members conducted topographic survey, geomorphological, glaciological, and geological studies, as well as named the area the Yamato Mountains after the ancient state of Yamato, the historical center of Japanese civilization (Asami et al., 1988).

In 1960s–1990s, the Belgica and Yamato (Queen Fabiola) Mountains were studied primarily by the JARE members (Yoshida and Fuiiwara, 1963; Kizaki, 1965; Tatsumi and Kizaki, 1969; Shiraishi et al., 1978; Kojima et al., 1981, 1982; Yanai et al., 1982; Yoshida, 1983; Asami et al., 1988; Motoyoshi et al., 1995). Some studies were also carried out by researchers of the Soviet Antarctic Expedition (Ravich et al., 1968) and Expedition Antarctique Belgo-Néerlandaise (Van Autenboer and Loy, 1972).

### *2.1 Physiography*

The Belgica Mountains (Figs. 2a and 3a) have a total area of about 33 km$^2$. They consist of three massifs—northwestern, southeastern, and southwestern—extending about 20 km from northeast to southwest. The northwestern massif includes Mounts Bastin, Maere, Gara, and Loodts as well as Kita, Naka, and Minami Ridges. The southeastern massif consists of Mounts Victor, Limburg Stirum, Solvay, Lahaye, and others. The southwestern massif incorporates Mounts Rossel, Kerckhove de Denterghem, Collard, and Paulus as well as Biôbu Ridge. Most of the mounts, rising 100–300 m above the ice surface, are characterized by steep slopes. There are several glacial troughs and cirques with glaciers (e.g., the Namako Glacier and Isidati Ice Fall) (Kojima et al., 1981, 1982). Elevation of most peaks ranges from 2,000 m to 2,500 m above sea level, with the highest peak of Mount Victor (2,590 m).

To the east of the southeastern massif, the icefield surface is approximately 300 m higher than to the west of the northwestern one (2,300–2,400 m vs. 1,950–2,050 m above sea level, respectively). There are two large glaciers in the Belgica Mountains (Figs. 2a and 3a): the Norsk Polarinstitutt Glacier located between the northwestern and southeastern massifs as well as the Giaever Glacier separated the southwestern and southeastern massifs. The moraine fields can be found mostly at northwestern slopes of the northwestern and southeastern massifs (Kojima et al., 1981, 1982).

The Yamato (Queen Fabiola) Mountains (Figs. 2b, 4, 5a, and 6a) consist of seven relatively small massifs, named from A to G, with a total area of about 115 km$^2$. They form a mountain chain extending about 50 km from north to south. The massifs rise 100–800 m from the ice sheet surface, which elevation ranges from 1,600 m to 2,100 m above sea level north and south of the mountains, respectively. Elevation of peaks ranges from 1,800 m to 2,400 m above sea level, with the highest peak of Mount Fukushima (2,470 m) in Massif D. The mountains have a predominantly alpine-type topography. The massifs are separated from each other by glaciers (Yoshida and Fuiiwara, 1963; Yoshida, 1983).

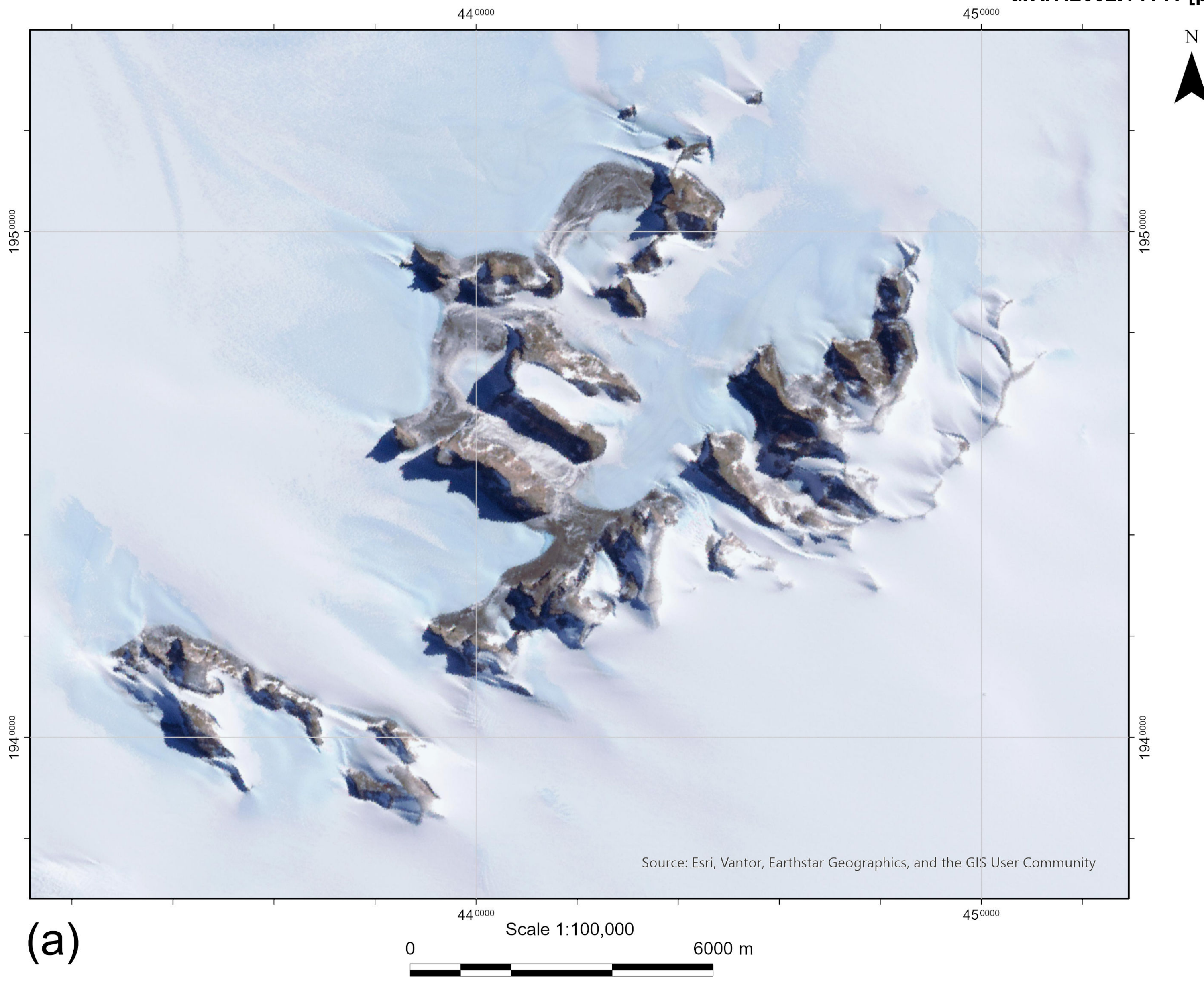

**Fig. 2** Satellite images of the study areas: (a) Belgica Mountains. Imagery is provided by Esri, Maxar, and Earthstar Geographics. Maps were created using ArcGIS Pro 3.0.1 (ESRI, 2015–2024). UTM projection, zone 36S for the Belgica Mountains and zone 37S for the Yamato (Queen Fabiola) Mountains.

*(Continued)*

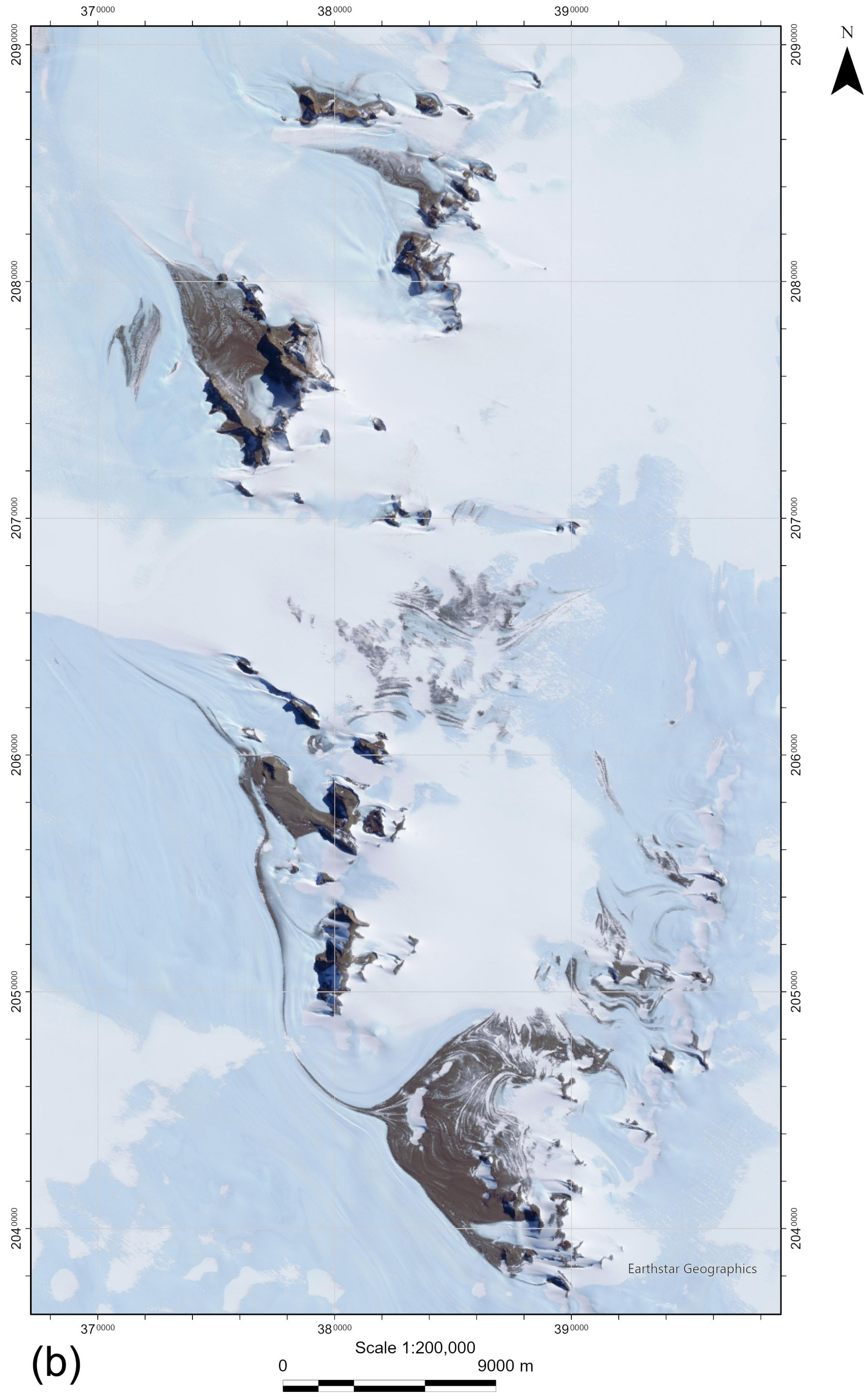

**Fig. 2, cont'd** Satellite images of the study areas: (b) Yamato (Queen Fabiola) Mountains.

The three northern massifs—Massifs G (Mounts De Breuck, Gyôten, and Van Pelt), F (Mount Pierre), and E (Mounts Goossens and Torimai)—have steep needle-like peaks with elevations ranging from 1,800 m to 2,200 m above sea level (Figs. 2b, 4, and 5a). The Torii Glacier, located to the southwest, separates these three massifs from Massif D.

Massif D (Figs. 2b, 4, and 5a) includes Mount Fukushima (2,470 m) and adjacent ridges with Kasuri Rock and Konsei Point, which form a northwest-opened glacial trough with sides parallel to the general direction of the ice sheet movement. The moraine field spreads in a fan-shaped concentric pattern from the trough onto the downstream ice sheet. The surface moraine has a wavy topography several meters high (Shiraishi et al., 1978).

Massif C (Figs. 2b, 4, and 6a) consists of several mounts and nunataks including Mounts Eyskens (2,300 m), Tyô (2,072 m), Sentyô (1,925 m), etc. It extends for about 13 km from northwest to southeast. The Yamato Glacier lies between Massif C and Massif D. Massif B (Figs. 2b, 4, and 6a), consisting of Mount Derom (2,400 m) and its spurs, extends for about 5 km from north to south. Massif B is separated from Massif C by the Christiaensen Glacier. Massifs C and B have steep northwestern and northern cliffs of glacial cirques and relatively gentle southeastern and eastern slopes, which are partially covered by moraine and snow (Kizaki, 1965; Yanai et al., 1982).

The southernmost Massif A (Figs. 2b, 4, and 6a) consists of Mount Gaston de Gerlanch (2,400 m) and several minor peaks and nunataks with gentle slopes. The Ôgi Glacier separates Massif A from Massif B. A chain of JARE-IV Nunataks extends 10 km north of Massif A trending from north-northeast to south-southwest. Most of the ice-free area is covered by extensive moraine deposits (Asami et al., 1988).

### *2.2 Geology*

The Belgica and Yamato (Queen Fabiola) Mountains constitute a late Neoproterozoic to Cambrian orogenic terrane of the Yamato–Belgica Complex. Its metamorphic and ultrametamorphic rocks belong to the crystalline basement of the Antarctic platform, which was formed in pre-Riphean times, but partially experienced early Paleozoic activation. The Yamato–Belgica Complex is characterized by low-pressure type metamorphism, c. 750 °C and less than 6 kbar as well as widespread igneous activity (Ravich and Kamenev, 1975; Hiroi et al., 1991; Satish-Kumar et al., 2008).

The Belgica Mountains (Figs. 2a and 3a) consist of crystalline basement rocks including granitic gneiss, hornblende-biotite banded gneiss, marble with skarn, amphibolite, and dyke rocks (i.e., basic-metadyke, syenite, granodiorite, diorite, and pink granite). These rocks form the Belgica Group, which is divided into the upper and lower formations. The Lower Belgica Formation is observed in the southwestern massif and the southwestern portion of the northwestern massif. This formation is dominated by granitic gneiss with amphibolite and marble and skarn beds. The Upper Belgica Formation is distributed in the main part of the northwestern massif and the southeastern massif. This formation includes alternation of hornblende-biotite melanocratic gneiss, biotite gneiss, and quartz-fildspathic gneiss with well-developed banded structure and amphibolite. The main part of the both massifs is occupied by hornblende-biotite melanocratic gneiss. Layers of crystalline limestone and thin beds of garnet-biotite gneiss, clinopyroxene gneiss, and augen gneiss are also observed. The age of the rocks of the Belgica Mountains ranges from 382 Ma to 472 Ma, suggesting early Paleozoic metamorphism (Kojima et al., 1981, 1982).

The Yamato (Queen Fabiola) Mountains are composed of metamorphic and plutonic rocks of the late Proterozoic and early Paleozoic ages. Two groups of metamorphic rocks can be distinguished. The first group is the granulite-facies rocks (syenite gneiss) observed in the northern and southern parts of the mountains. The second group is the amphibolite-facies rocks (granite gneiss and migmatite gneiss) occupying the central part of the mountains.

Granulite-facies and amphibolite-facies metamorphisms are associated with two events at ca. 620 Ma and ca. 535 Ma, respectively. Syenites, the main type of local plutonic rocks, were intruded into the metamorphic rocks (Ravich and Kamenev, 1975; Shiraishi et al., 1978; Yanai et al., 1982; Asami et al., 1988; Ruppel et al., 2018).

## 3 Materials and methods

We used two digital elevation models (DEMs) with a grid spacing of 10 m for the Belgica Mountains and 32 m for the Yamato (Queen Fabiola) Mountains. The DEMs were extracted from the Reference Elevation Model of Antarctica (REMA) (Howat et al., 2019; PGC, 2018–2024). For the Belgica Mountains, the 20-m gridded DEM was obtained by resampling a 10-m gridded DEM. For data processing and mapping, we applied a procedure described earlier in our papers (Florinsky, 2025b; Florinsky and Zharnova, 2025b, 2025c, 2025d, 2025e).

### *3.1 Preprocessing*

REMA is presented in the polar stereographic projection with an elevation datum of the WGS84 ellipsoid. The extracted DEMs were reprojected into the UTM projection, zone 36S for the Belgica Mountains and zone 37S for the Yamato (Queen Fabiola) Mountains. The original grid spacings were preserved. Ellipsoidal elevations were transformed into orthometric ones.

For the Yamato (Queen Fabiola) Mountains region, REMA includes high-frequency noise within some glacier areas. This noise causes artifact patterns on morphometric maps. To suppress the noise, we applied double Gaussian filtering to the Yamato (Queen Fabiola) DEM.

### *3.2 Calculations*

Digital models of eleven, most scientifically important morphometric variables were derived from the reprojected DEMs. The list of morphometric variables (Table 2) includes six local attributes: slope (*G*), aspect (*A*), horizontal curvature ($k_h$), vertical curvature ($k_v$), minimal curvature ($k_{min}$), and maximal curvature ($k_{max}$); one nonlocal variable—catchment area (*CA*); two combined variables: topographic wetness index (*TWI*) and stream power index (*SPI*); as well as two two-field-specific attributes—total insolation (*TIns*) and wind exposition index (*WEx*). Formulas and detailed interpretations of these morphometric variables can be found elsewhere (Shary et al., 2002; Florinsky 2017, 2025a, chap. 2).

To derive digital models of local variables, we used a finite-difference method by Evans (1980). To compute models of *CA*, we applied a maximum-gradient based multiple flow direction algorithm by Qin et al. (2007) to preprocessed sink-filled DEMs. *CA* digital models were then logarithmized. To derive digital models of combined morphometric variables, we used calculated models of *CA* and *G*. To compute digital models of two-field-specific attributes, we applied two related methods by Böhner (2004). *TIns* were estimated for one mid-summer day (1st January) with a temporal step of 0.5 h.

### *3.3 Mapping*

The Yamato (Queen Fabiola) Mountains are quite large to represent in detail their morphometric characteristics on one map sheet. In this regard, we divided its digital morphometric models into two parts and mapped them on separate sheets. The sheet layout can be found on the overview hypsometric map of the Yamato (Queen Fabiola) Mountains (Fig. 4).

**Table 2** Definitions and interpretations of key morphometric variables (Shary et al., 2002; Florinsky, 2017, 2025a, chap. 2).

| Variable, notation, and unit | Definition and interpretation |
|---|---|
| | *Local morphometric variables* |
| Slope, *G* (°) | An angle between the tangential and horizontal planes at a given point of the topographic surface. Relates to the velocity of gravity-driven flows. |
| Aspect, *A* (°) | An angle between the north direction and the horizontal projection of the two-dimensional vector of gradient counted clockwise, from 0° to 360°, at a given point of the topographic surface. Relates to the direction of gravity-driven flows |
| Horizontal (tangential) curvature, $k_h$ ($m^{-1}$) | A curvature of a normal section tangential to a contour line at a given point of the surface. A measure of flow convergence and divergence. Gravity-driven lateral flows converge where $k_h < 0$, and diverge where $k_h > 0$. $k_h$ mapping reveals crest and valley spurs. |
| Vertical (profile) curvature, $k_v$ ($m^{-1}$) | A curvature of a normal section having a common tangent line with a slope line at a given point of the surface. A measure of relative deceleration and acceleration of gravity-driven flows. They are decelerated where $k_v < 0$, and are accelerated where $k_v > 0$. $k_v$ mapping reveals terraces and scarps. |
| Minimal curvature, $k_{min}$ ($m^{-1}$) | A curvature of a principal section with the lowest value of curvature at a given point of the surface. $k_{min} > 0$ corresponds to local convex landforms, while $k_{min} < 0$ relates to elongated concave landforms (e.g., hills and troughs, correspondingly). |
| Maximal curvature, $k_{max}$ ($m^{-1}$) | A curvature of a principal section with the highest value of curvature at a given point of the surface. $k_{max} > 0$ corresponds to elongated convex landforms, while $k_{max} < 0$ relate to local concave landforms (e.g., crests and holes, correspondingly). |
| | *Nonlocal morphometric variables* |
| Catchment area, *CA* ($m^2$) | An area of a closed figure formed by a contour segment at a given point of the surface and two flow lines coming from upslope to the contour segment ends. A measure of the contributing area. |
| | *Combined morphometric variables* |
| Topographic wetness index, *TWI* | A ratio of catchment area to slope gradient at a given point of the topographic surface. A measure of the extent of flow accumulation. |
| Stream power index, *SPI* | A product of catchment area and slope gradient at a given point of the topographic surface. A measure of potential flow erosion and related landscape processes. |
| | *Two-field specific morphometric variables* |
| Total insolation, *TIns* (kWh/$m^2$) | A measure of the topographic surface illumination by solar light flux. Total potential incoming solar radiation, a sum of direct and diffuse insolations. |
| Wind exposition index, *WEx* | A measure of an average exposition of slopes to wind flows of all possible directions at a given point of the topographic surface. |

First, we created hypsometric maps of the Belgica Mountains and the Yamato (Queen Fabiola) Mountains from the reprojected DEMs. Two gradient hypsometric tint scales were used. To display the elevations of the ice-free topography, we applied a yellow-brown part of the standard spectral hypsometric scale of color plasticity (Kovaleva, 2014). To display the elevations of the glacier topography, we utilized a portion of a modified hypsometric tint scale for polar regions (Patterson and Jenny, 2011). Hypsometric tintings were then combined with achromatic hill shading derived from the DEMs by a standard procedure (Figs. 3a, 4, 5a, and 6a).

Next, we placed labels of geographical names on the hypsometric maps. As a source of this information, we used maps of the study areas (Tatsumi and Kizaki, 1969; Shiraishi et al., 1978; Kojima et al., 1981; Yanai et al., 1982; Asami et al., 1988; Motoyoshi et al., 1995; GIAJ, 2016, 2018) and the Composite Gazetteer of Antarctica (SCAR, 2000–2014). We utilized the following principles:

- Japanese maps utilize two romanization systems for transcribing Japanese words into the Latin alphabet (i.e., either the Kunrei-shiki system or the Hepburn one). The Kunrei-shiki system is, as a rule, used in the Composite Gazetteer of Antarctica (SCAR, 2000–2014). We used transcriptions of the Japanese geographic names from the Composite Gazetteer.
- Specific proper names were used without modification. However, we replaced generic names of Japanese origin (e.g., dai, daira, dake, hyoga, iwa, kubo, nagaone, san, yama, etc) with English ones (e.g., cirque, glacier, heights, icefield, hills, mount, mountains, peak, ridge, rock, valley, etc.) for consistency and avoiding tautologies.

Second, from the calculated digital morphometric models, we produced three series of morphometric maps for the Belgica Mountains and the Yamato (Queen Fabiola) Mountains (Figs. 3b–l, 5b–l, and 6b–l). For optimal visual perception of morphometry, we used the following rules for applying gradient tint scales:

1. *G* and *CA* can take only positive values. To map *G* and *CA*, we applied a standard gray tint scale; the minimum and maximum values of *G* or *CA* correspond to white and black, respectively (Figs. 3b and 3h, 5b and 5h, 6b and 6h).
2. $k_h$, $k_v$, $k_{min}$, and $k_{max}$ can take both negative and positive values, having opposite physical mathematical and physical geographical sense and interpretation. To map curvatures, we used a two-color tint scale consisting of two contrasting parts, blue and orange (negative and positive values, respectively). The most and least saturated shades of blue or orange colors correspond to the absolute maximum and absolute minimum values, respectively, of $k_h$, $k_v$, $k_{min}$, and $k_{max}$ (Figs. 3d–g, 5d–g, 6d–g).
3. *TWI* and *SPI* can take only positive values. To map these indices, we applied a standard spectral tint scale; the minimum and maximum values of *TWI* or *SPI* correspond to violet and red, respectively (Figs. 3i and 3j, 5i and 5j, 6i and 6j).
4. *TIns* is a nonnegative variable. To map it, we used an orange tint scale: the minimum and maximum *TIns* values correspond to the darkest and lightest orange shades depicting the least and most illuminated areas, respectively (Figs. 3k, 5k, and 6k).
5. *WEx* is a positive dimensionless variable, wherein values below and above 1 relate to wind-shadowed and -exposed areas, respectively. To map this index, we applied a two-color tint scale consisting of two contrasting parts, orange and violet (values below and above 1, respectively). The darkest shades of orange and violet colors correspond to the minimum and maximum *WEx* values, respectively, while the lightest shades of the colors correspond to 1 (Figs. 3l, 5l, and 6l).

All maps were produced in 1:100,000 scale, with UTM projection, zone 36S for the Belgica Mountains and zone 37S for the Yamato (Queen Fabiola) Mountains.

For DEM processing and geomorphometric calculations, we used a software SAGA 9.8.1 (Conrad et al., 2015). For morphometric mapping, we utilized ArcGIS Pro 3.0.1 (ESRI, 2015–2024).

## 4 Results

Geomorphometric modeling and mapping resulted in the series of morphometric maps for two, partly ice-free mountainous areas of the eastern Queen Maud Land including the Belgica Mountains (Fig. 3) and the Yamato (Queen Fabiola) Mountains (Figs. 5 and 6).

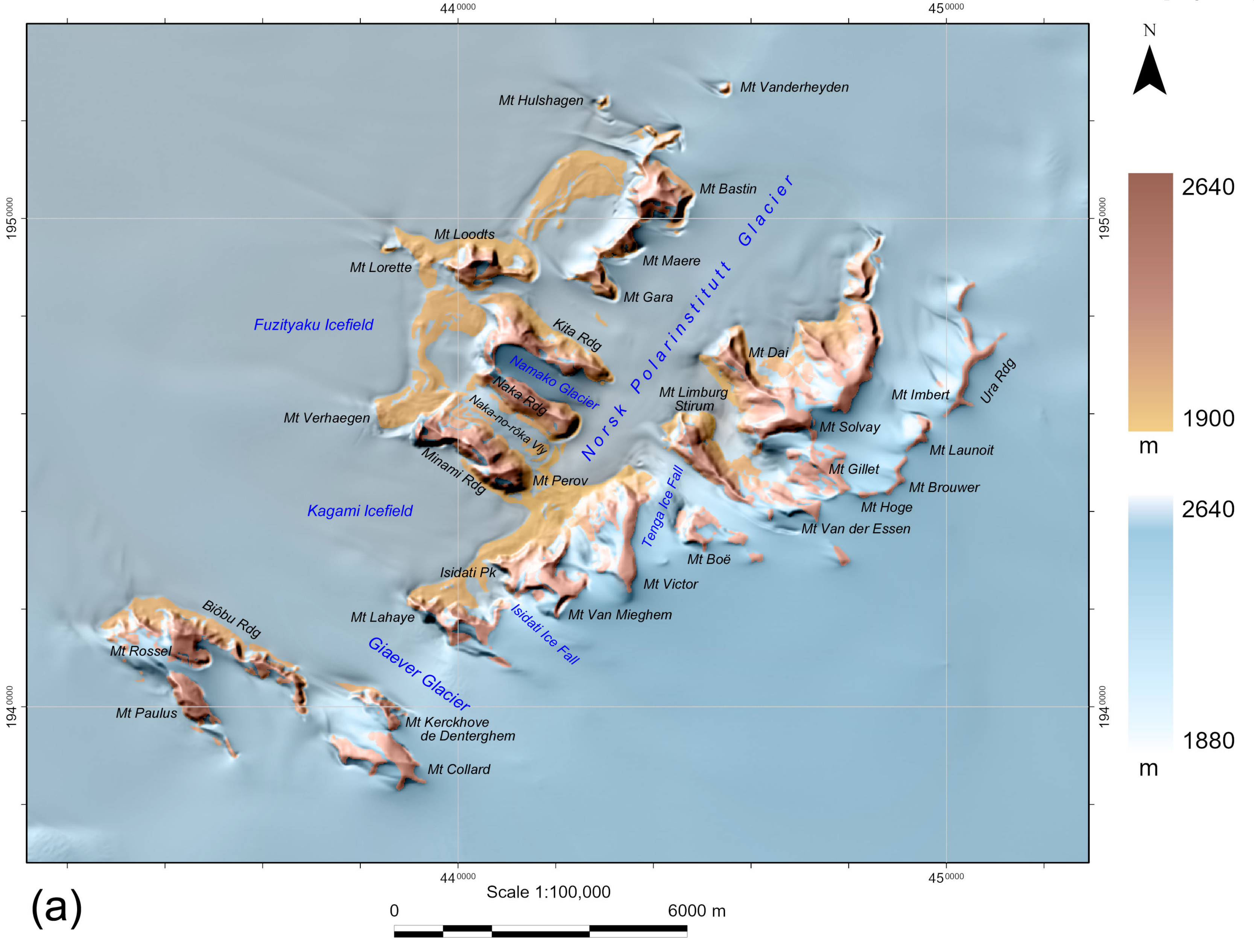

**Fig. 3** Belgica Mountains: (a) Elevation.

*(Continued)*

**Fig. 3, cont'd** Belgica Mountains: (b) Slope.

*(Continued)*

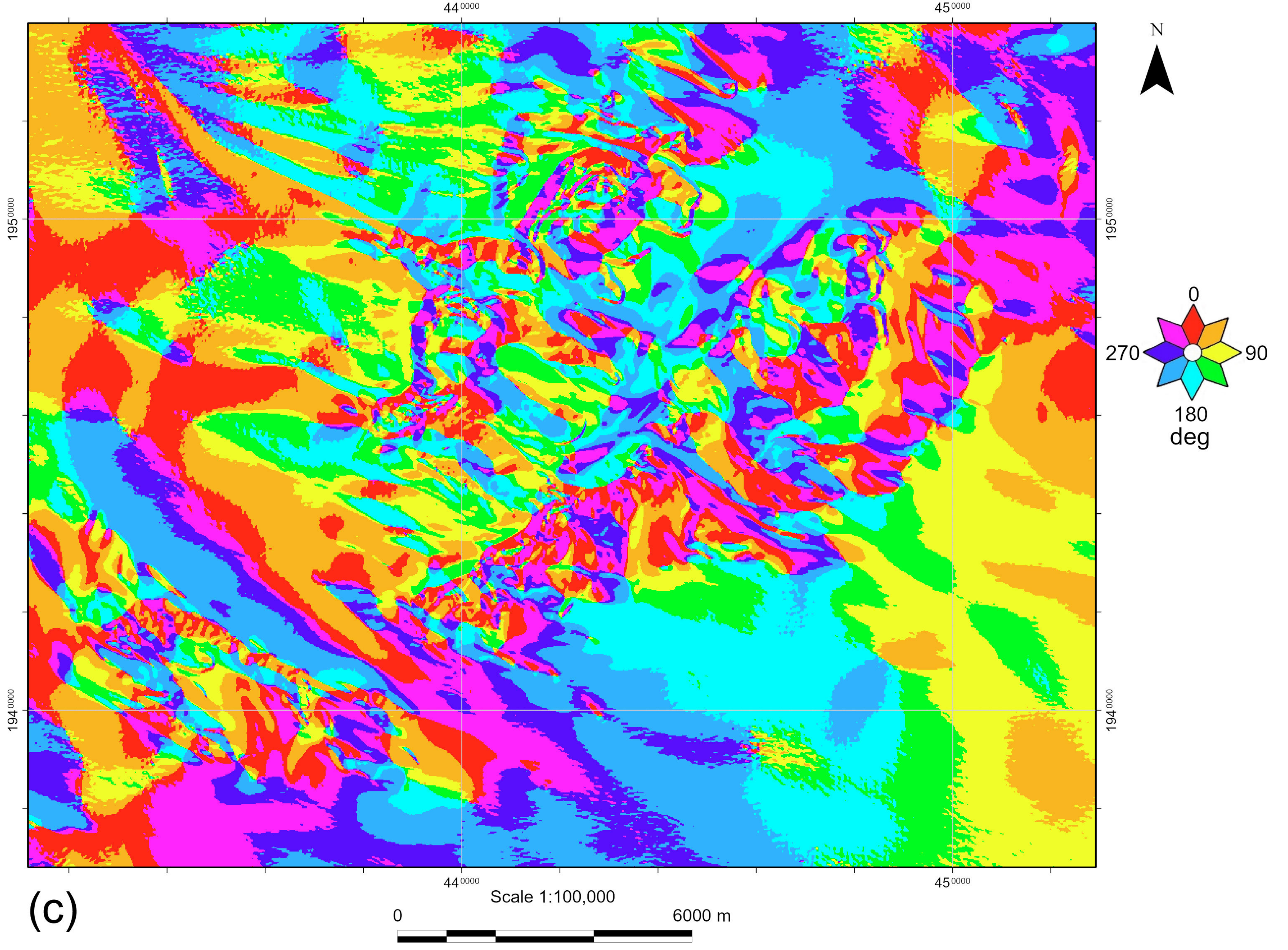

**Fig. 3, cont'd** Belgica Mountains: (c) Aspect.

*(Continued)*

(d)

Scale 1:100,000

0 6000 m

**Fig. 3, cont'd** Belgica Mountains: (d) Horizontal curvature.

*(Continued)*

(e)

Scale 1:100,000

0 6000 m

**Fig. 3, cont'd** Belgica Mountains: (e) Vertical curvature.

*(Continued)*

(f)

Scale 1:100,000

0 6000 m

**Fig. 3, cont'd** Belgica Mountains: (f) Minimal curvature.

*(Continued)*

**Fig. 3, cont'd** Belgica Mountains: (g) Maximal curvature.

*(Continued)*

N

17
6
log
scale

(h)

Scale 1:100,000
0 6000 m

**Fig. 3, cont'd** Belgica Mountains: (h) Catchment area.

*(Continued)*

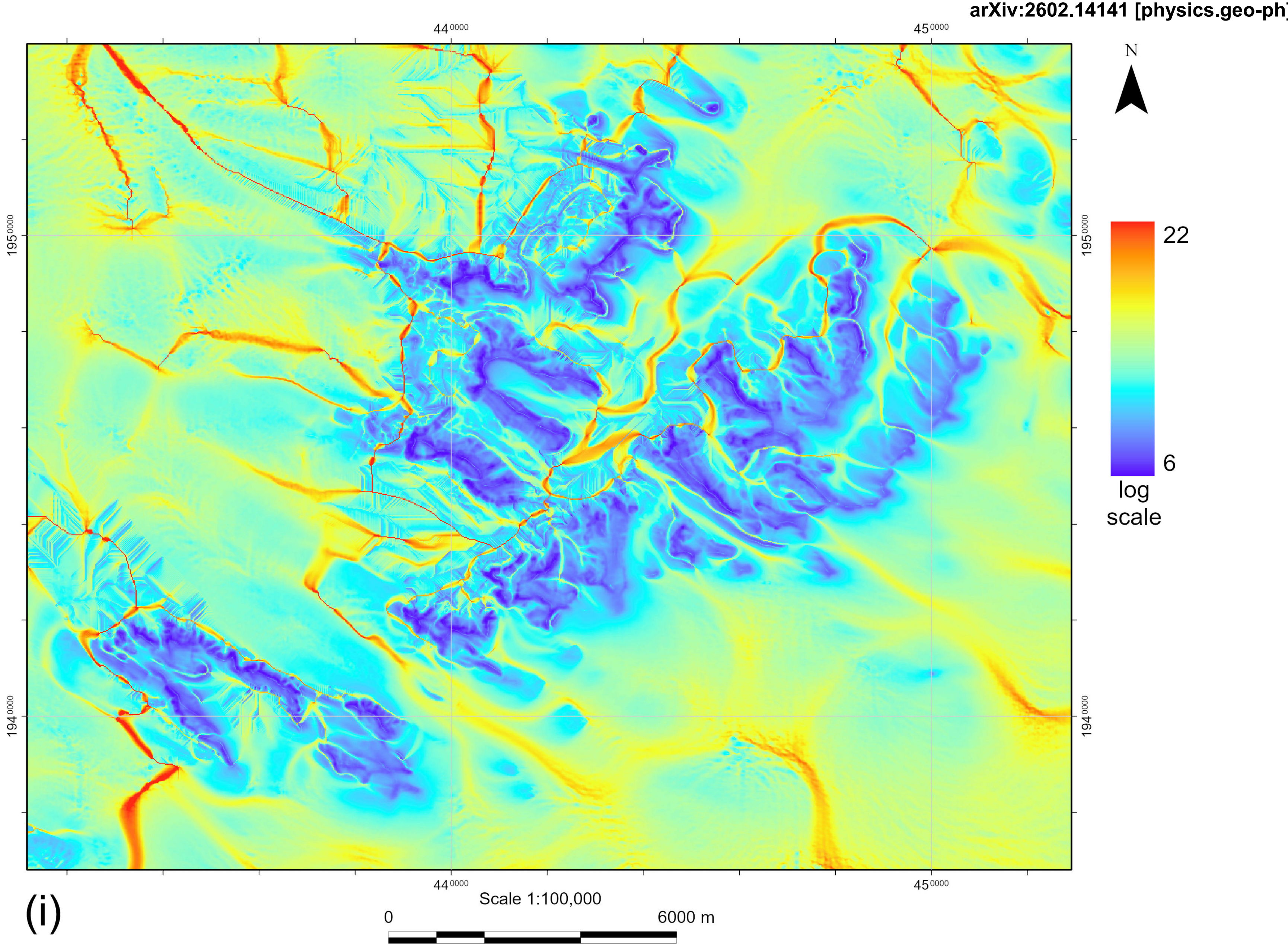

**Fig. 3, cont'd** Belgica Mountains: (i) Topographic wetness index.

*(Continued)*

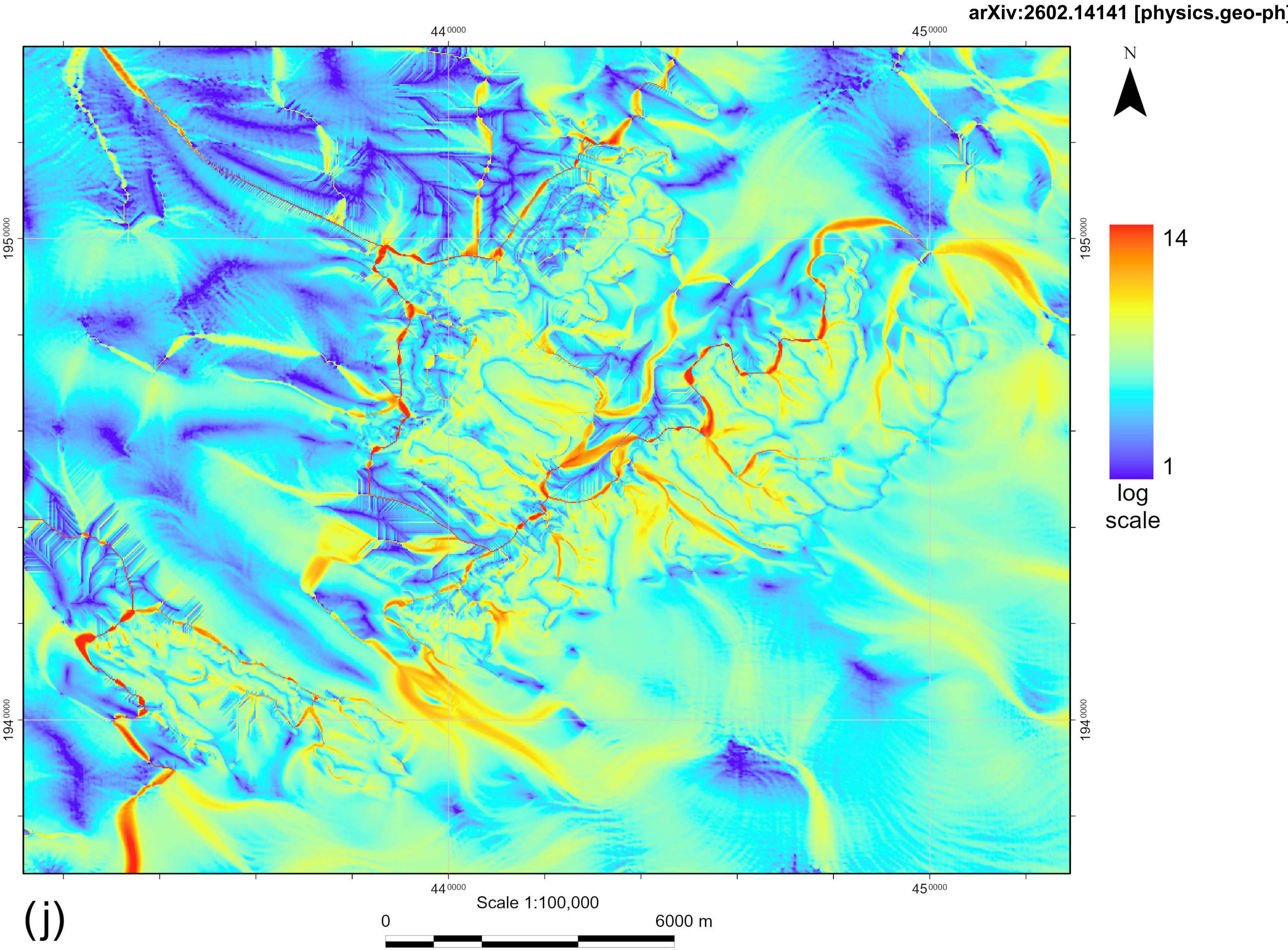

**Fig. 3, cont'd** Belgica Mountains: (j) Stream power index.

*(Continued)*

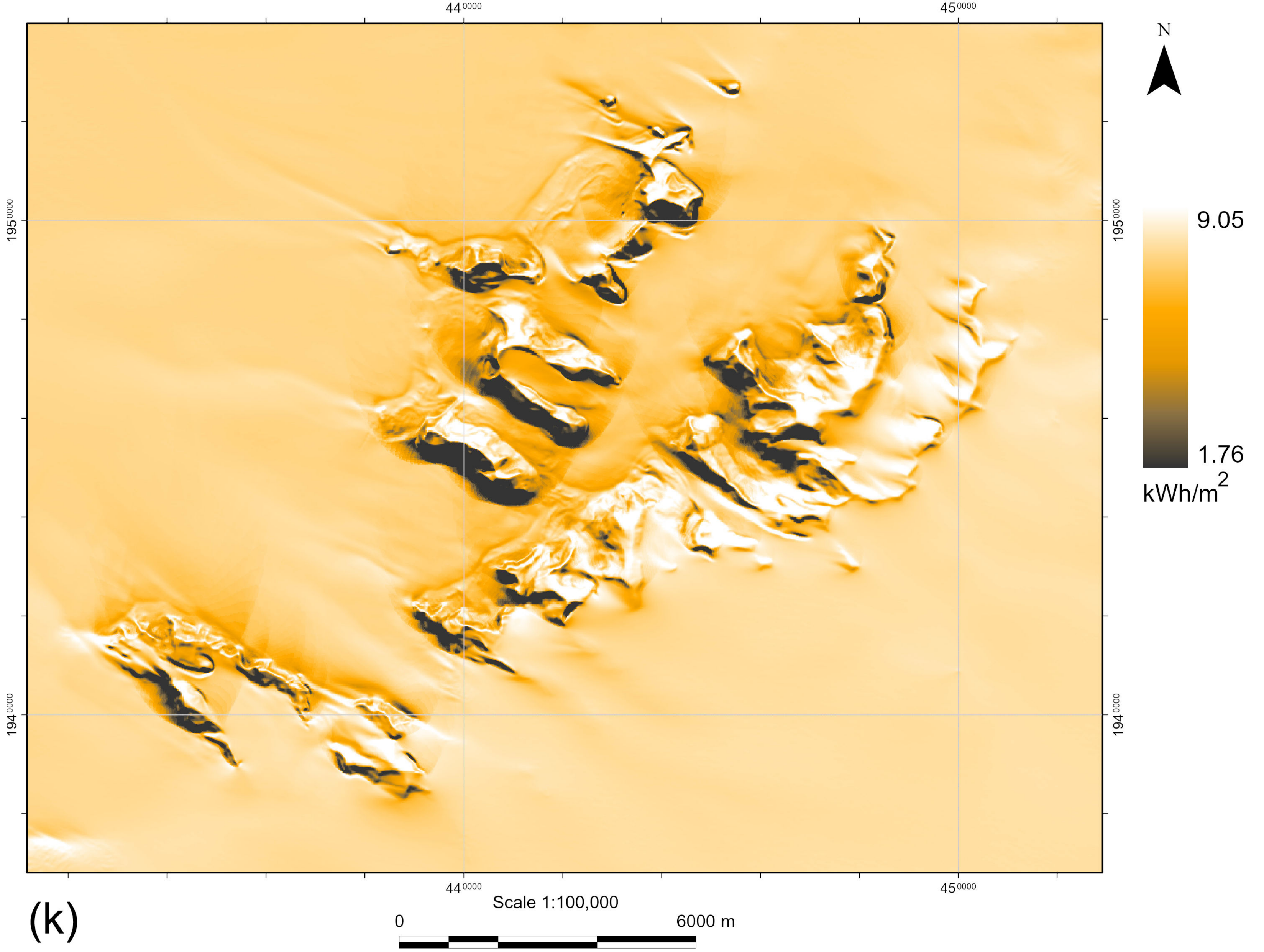

**Fig. 3, cont'd** Belgica Mountains: (k) Total insolation.

*(Continued)*

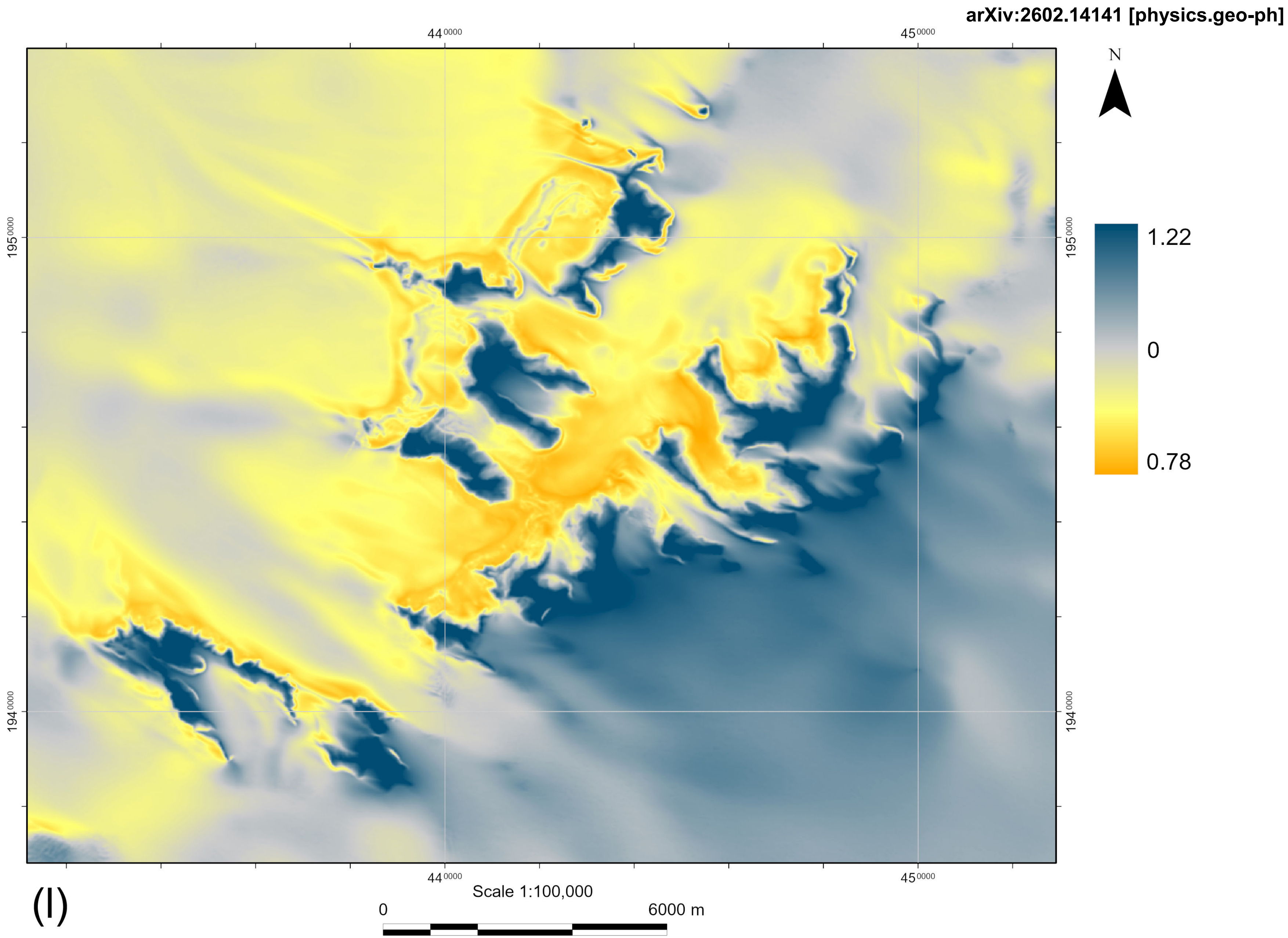

**Fig. 3, cont'd** Belgica Mountains: (l) Wind exposition index.

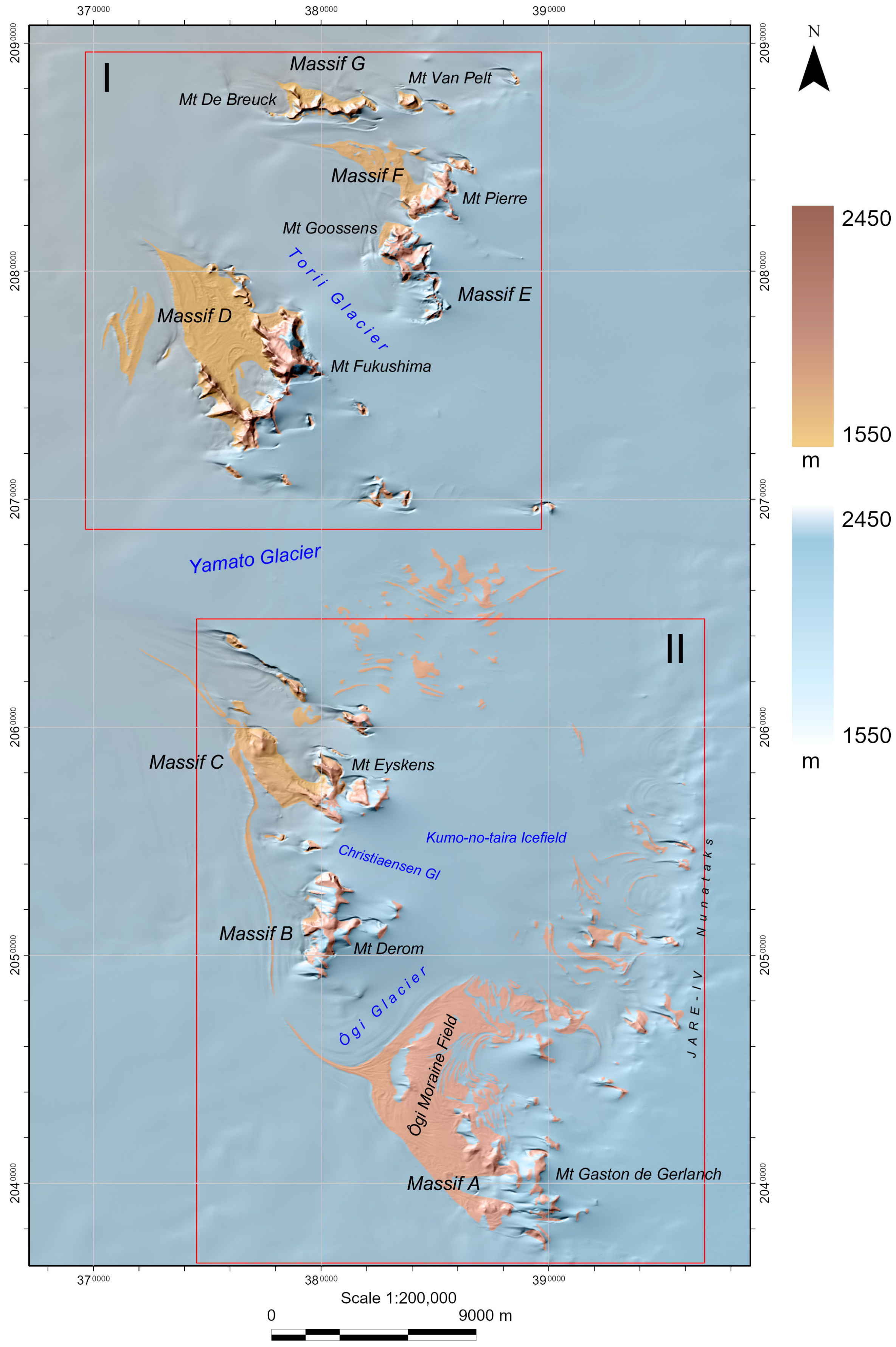

**Fig. 4** Yamato (Queen Fabiola) Mountains: an overview hypsometric map with a layout of sheets (red rectangles with Latin numerals).

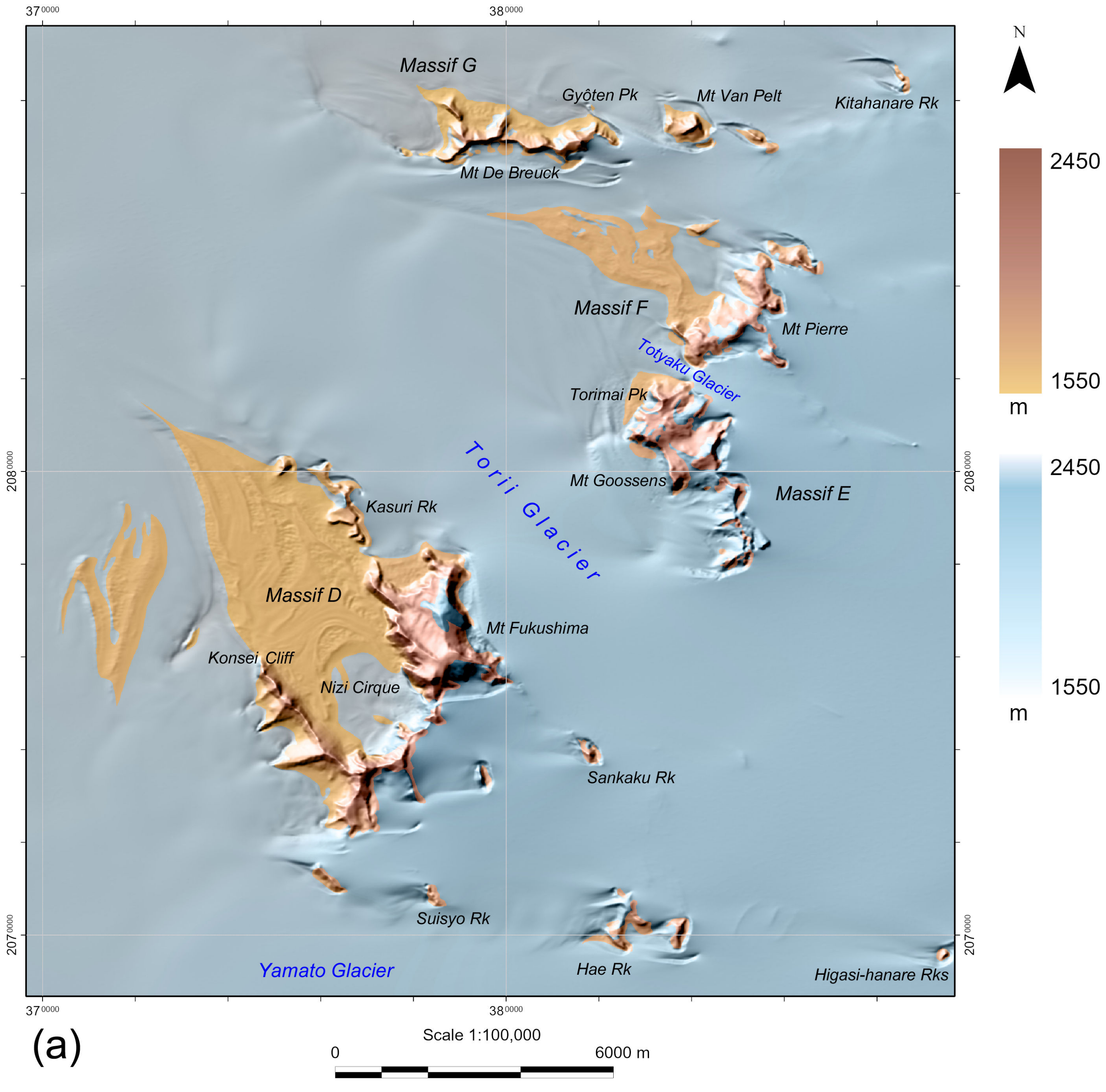

**Fig. 5** Yamato (Queen Fabiola) Mountains, sheet I: (a) Elevation.

*(Continued)*

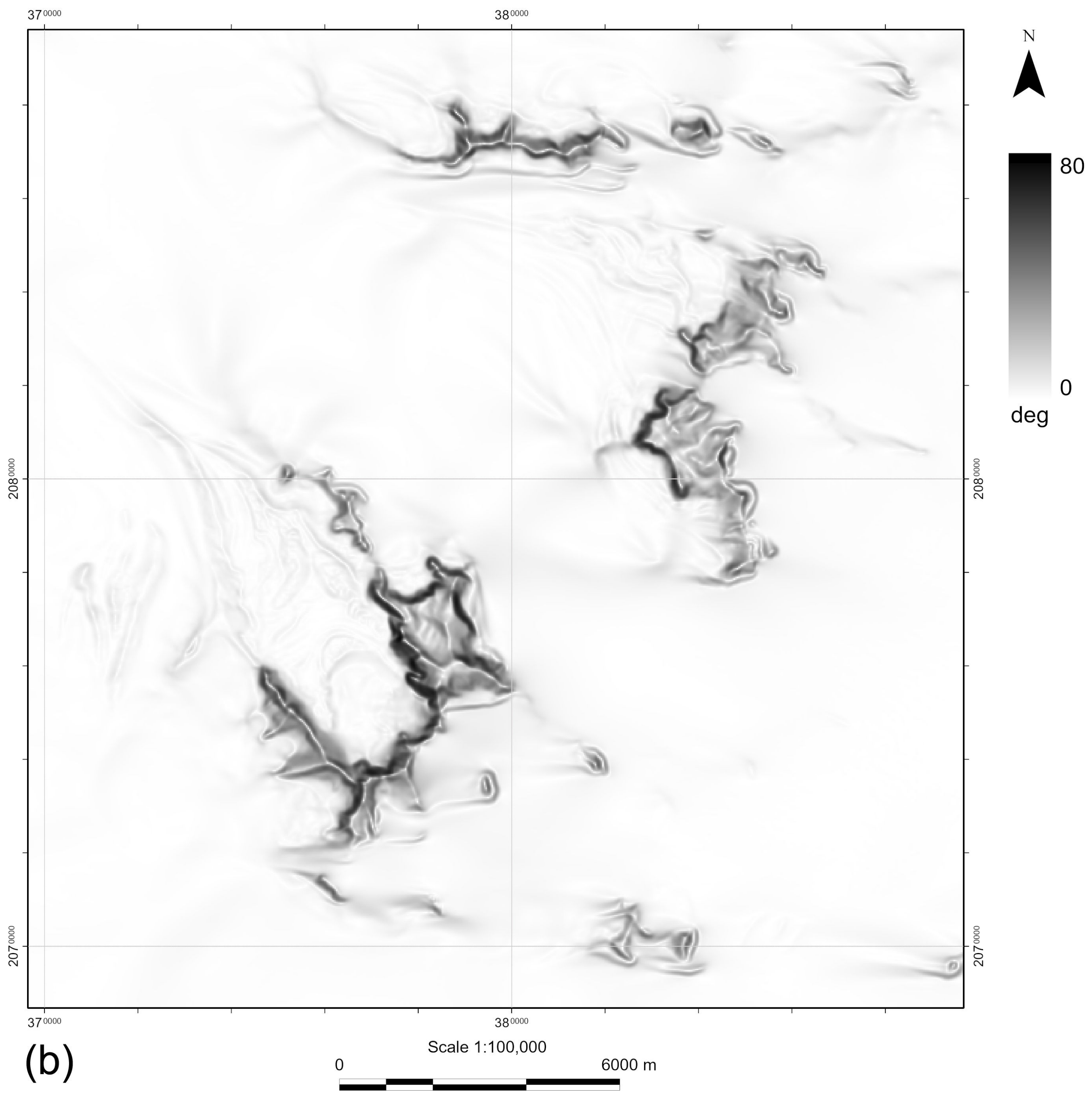

**Fig. 5, cont'd** Yamato (Queen Fabiola) Mountains, sheet I: (b) Slope.

*(Continued)*

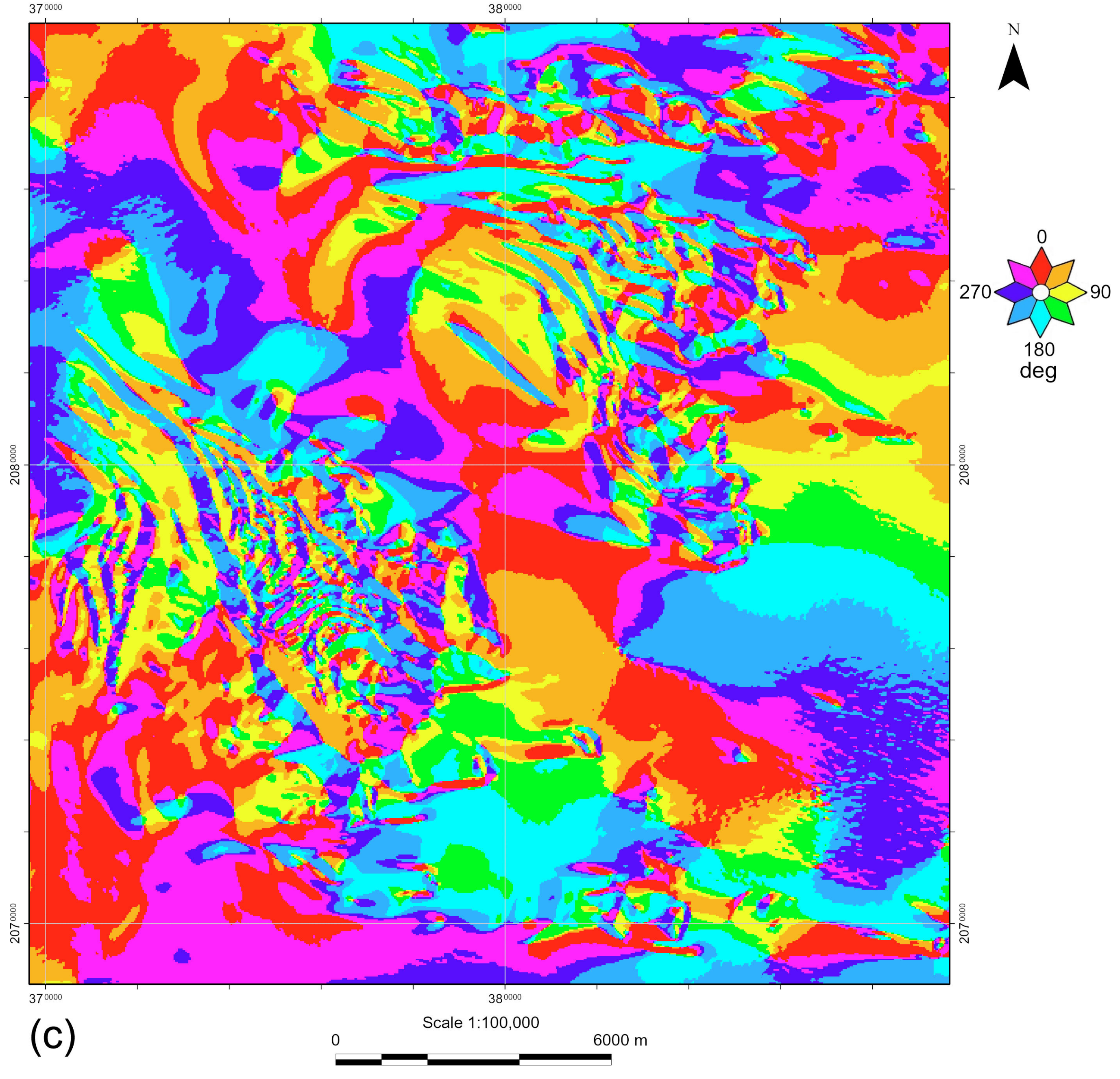

**Fig. 5, cont'd** Yamato (Queen Fabiola) Mountains, sheet I: (c) Aspect.

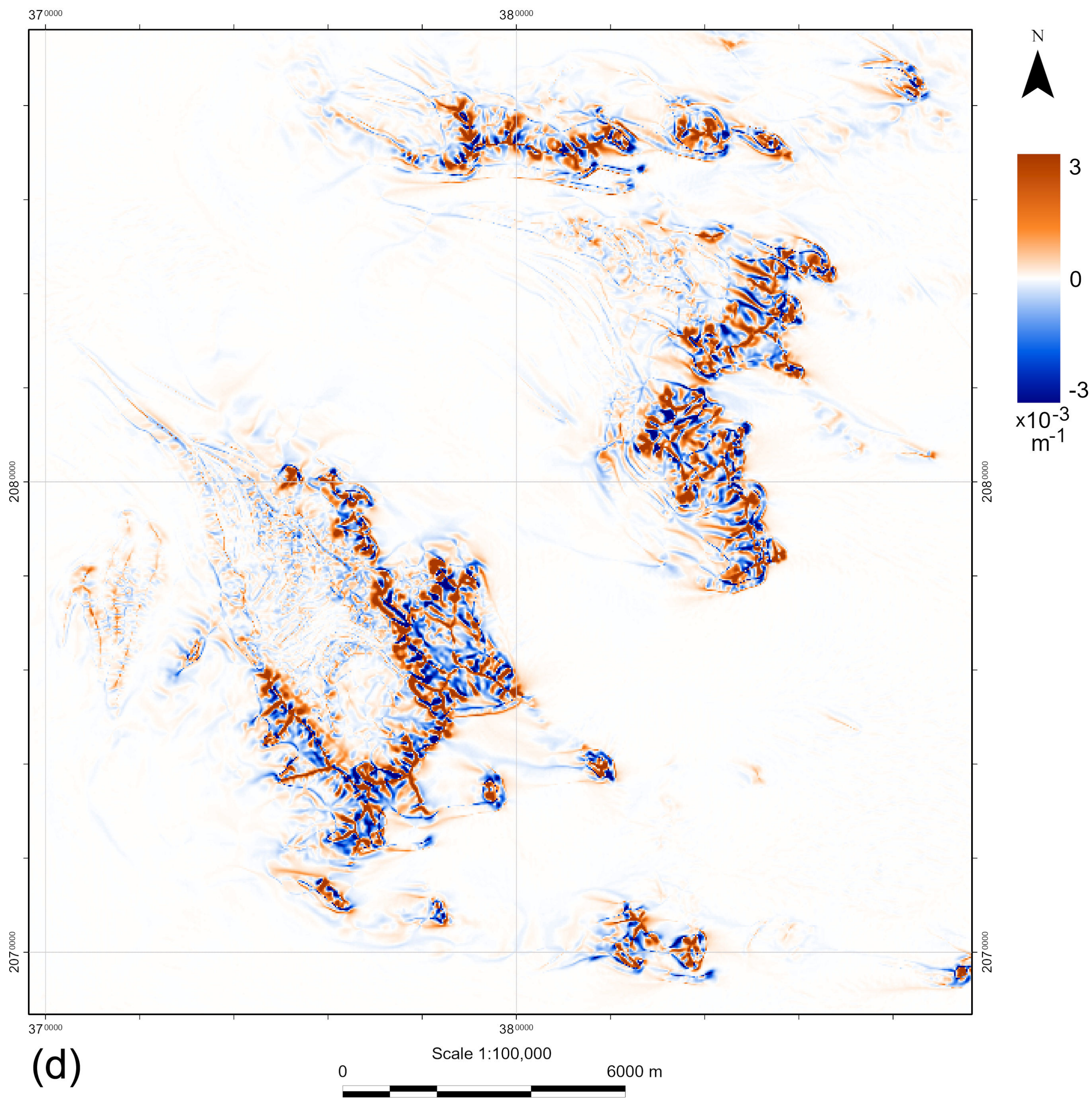

**Fig. 5, cont'd** Yamato (Queen Fabiola) Mountains, sheet I: (d) Horizontal curvature.

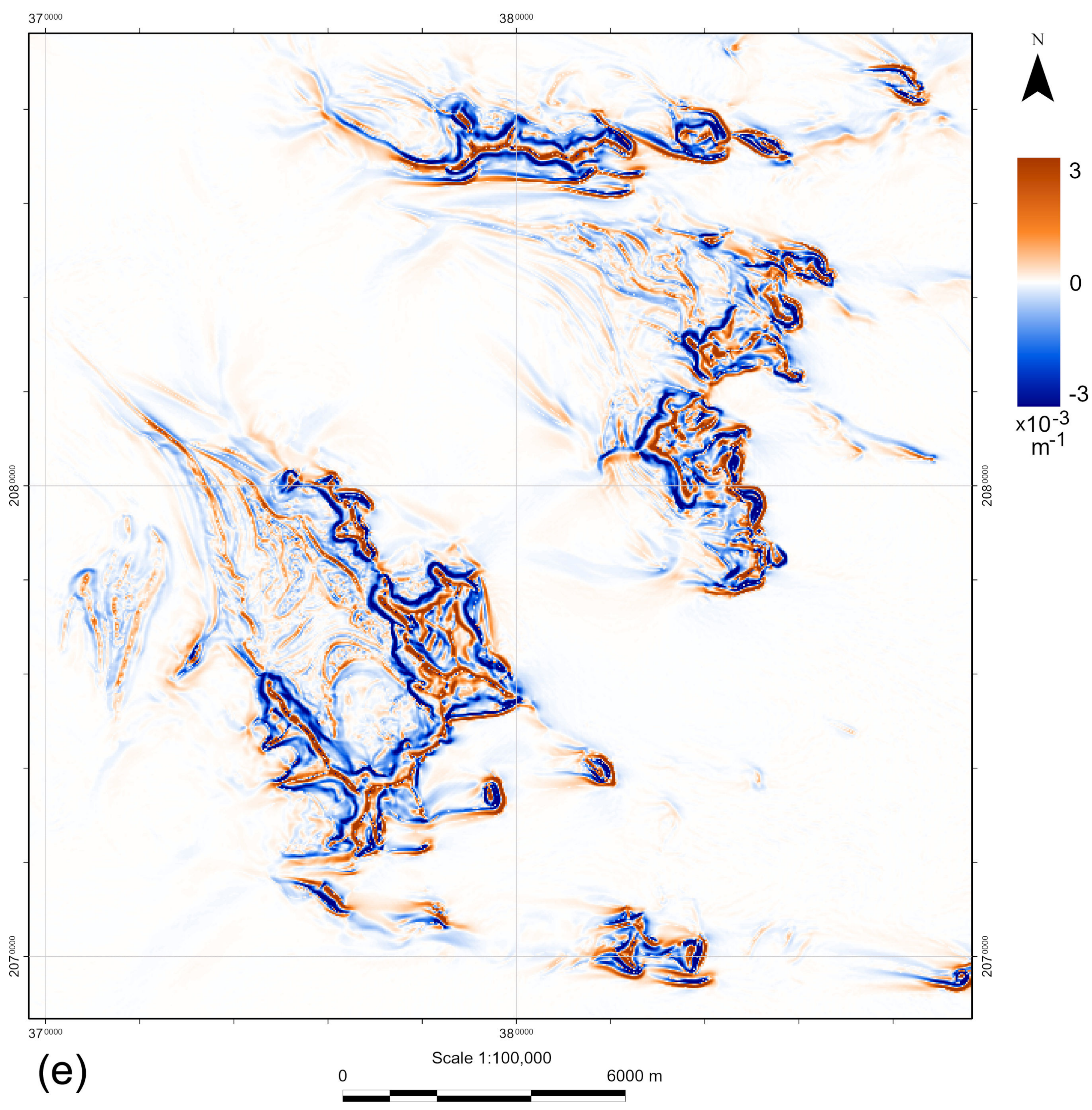

**Fig. 5, cont'd** Yamato (Queen Fabiola) Mountains, sheet I: (e) Vertical curvature.

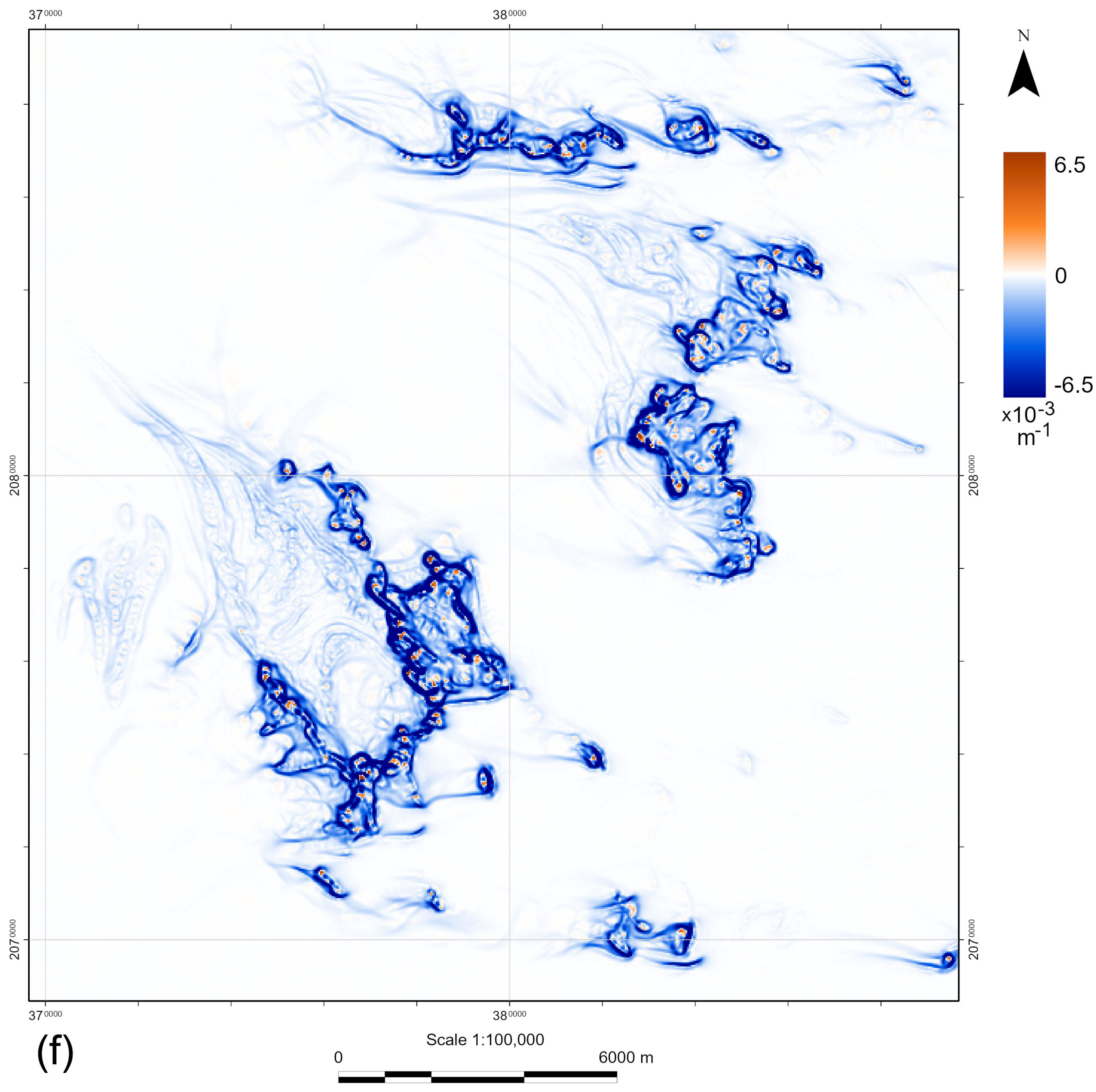

**Fig. 5, cont'd** Yamato (Queen Fabiola) Mountains, sheet I: (f) Minimal curvature.

*(Continued)*

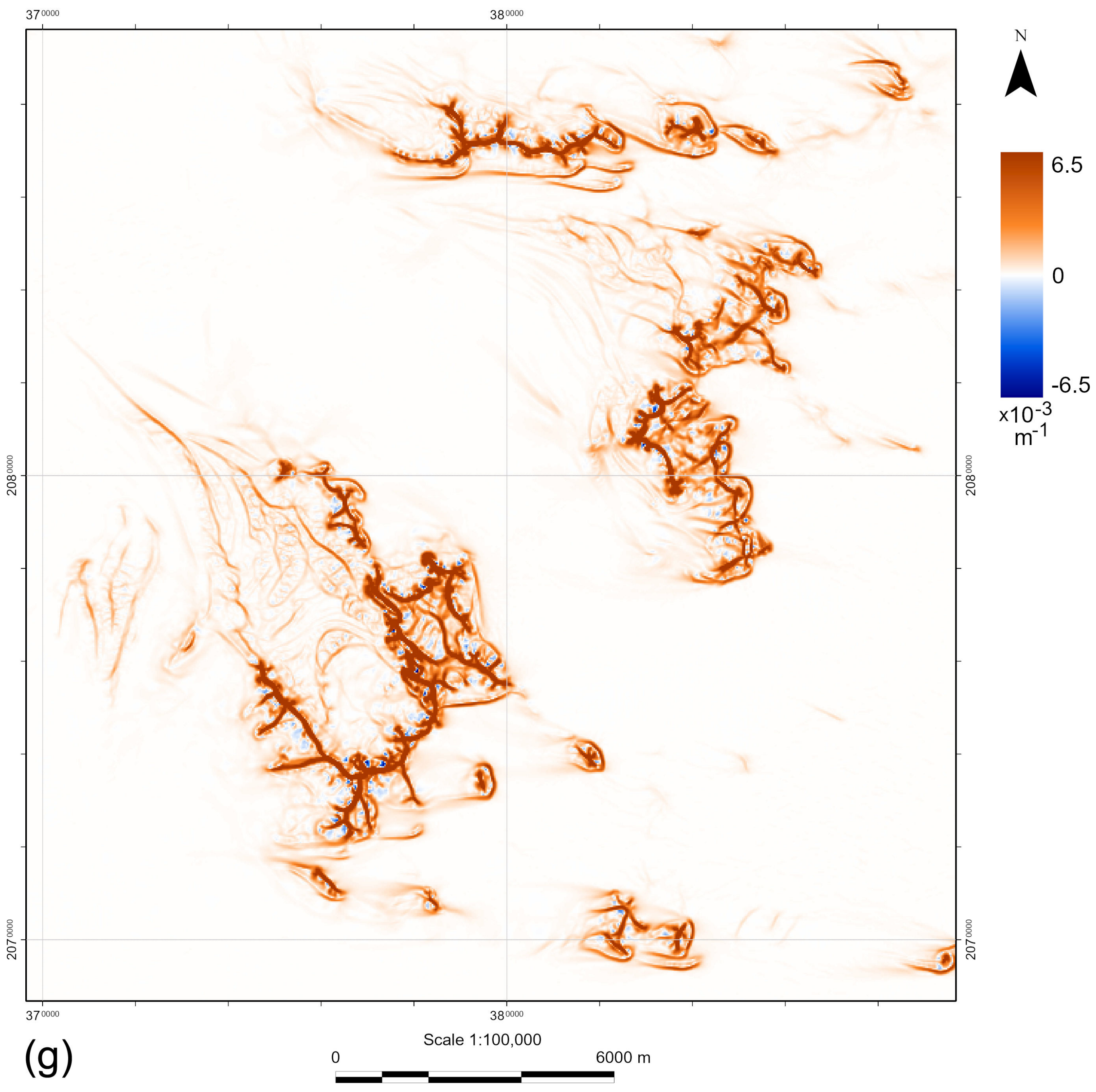

**Fig. 5, cont'd** Yamato (Queen Fabiola) Mountains, sheet I: (g) Maximal curvature.

*(Continued)*

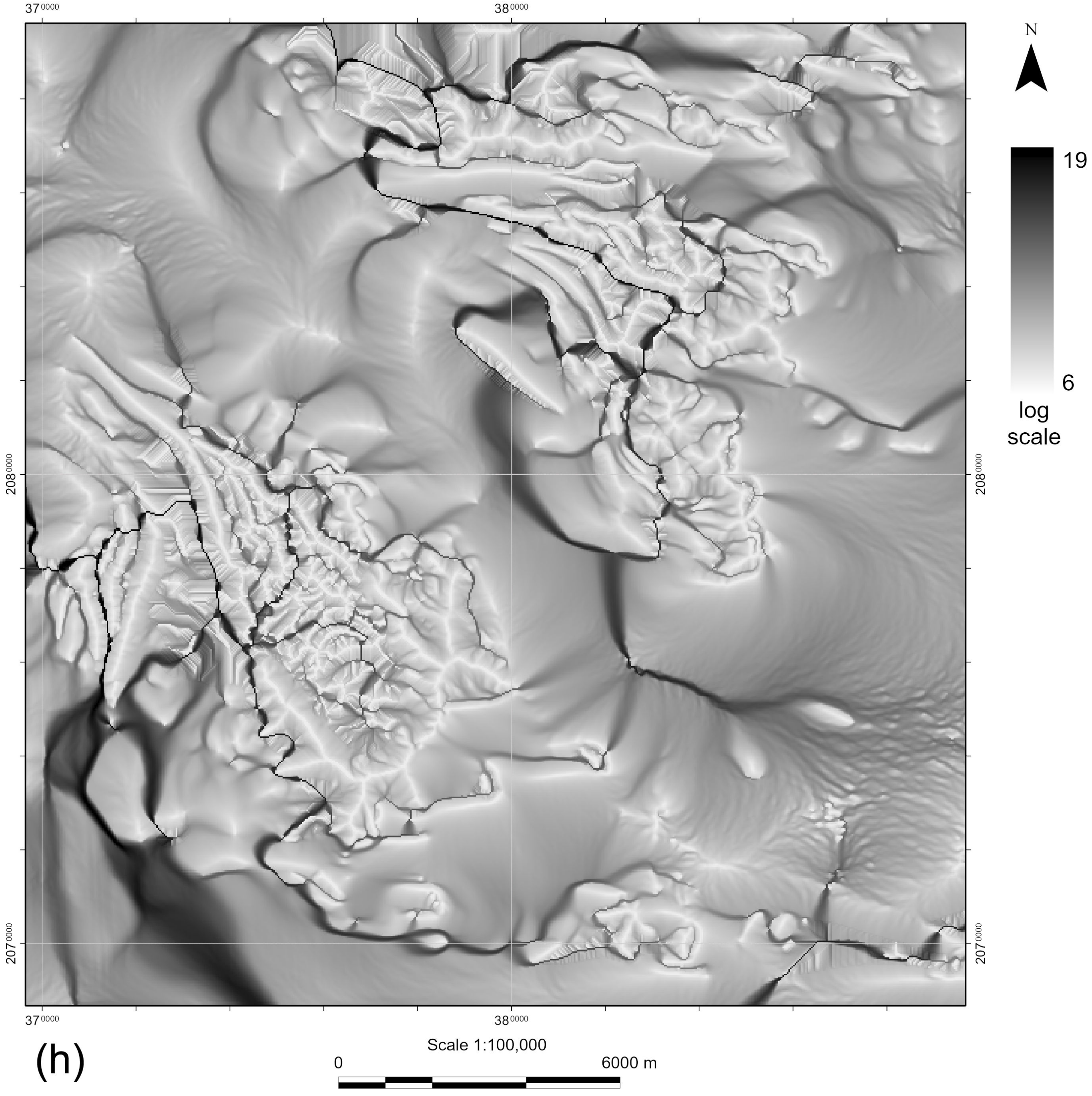

**Fig. 5, cont'd** Yamato (Queen Fabiola) Mountains, sheet I: (h) Catchment area.

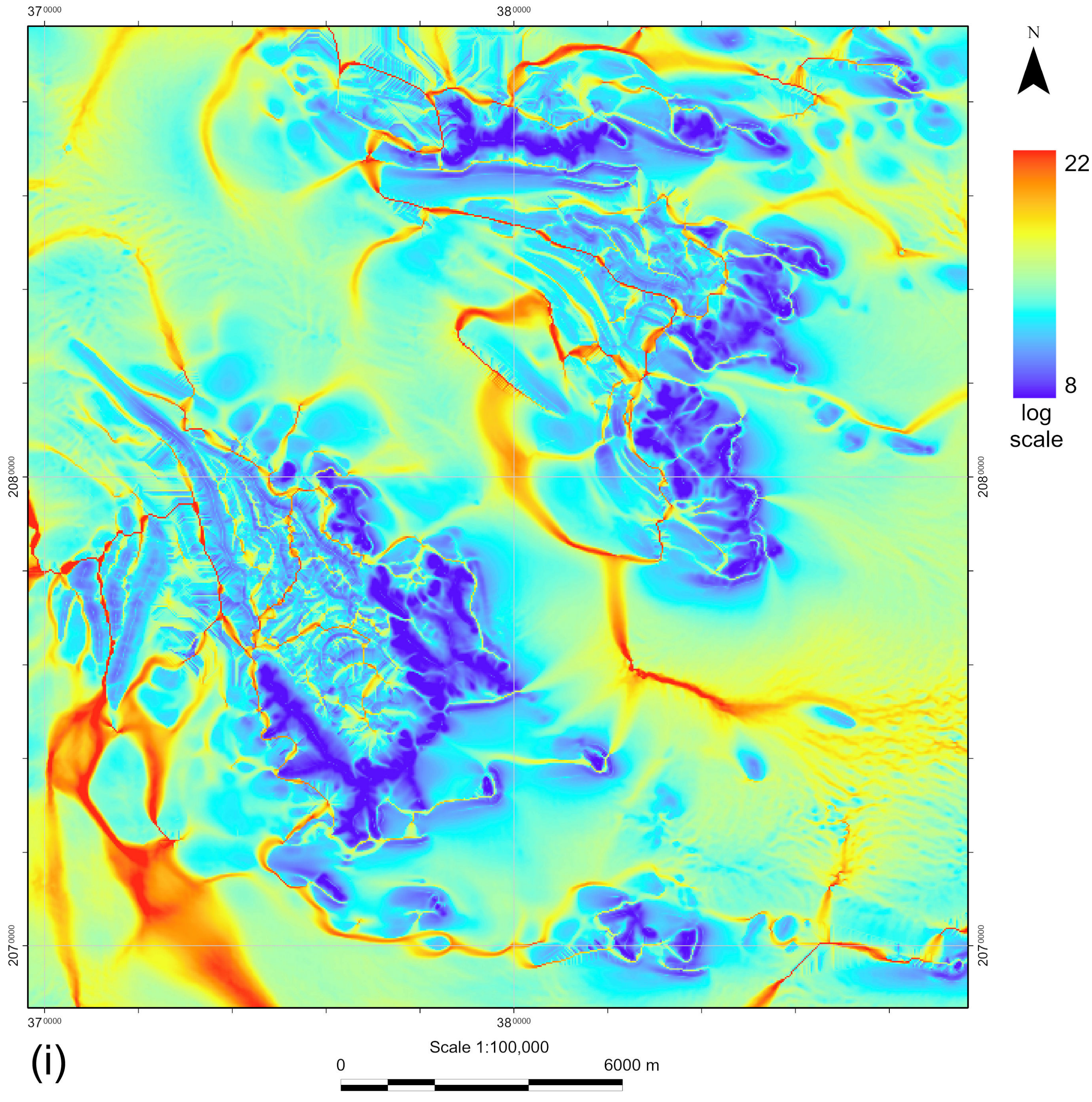

**Fig. 5, cont'd** Yamato (Queen Fabiola) Mountains, sheet I: (i) Topographic wetness index.

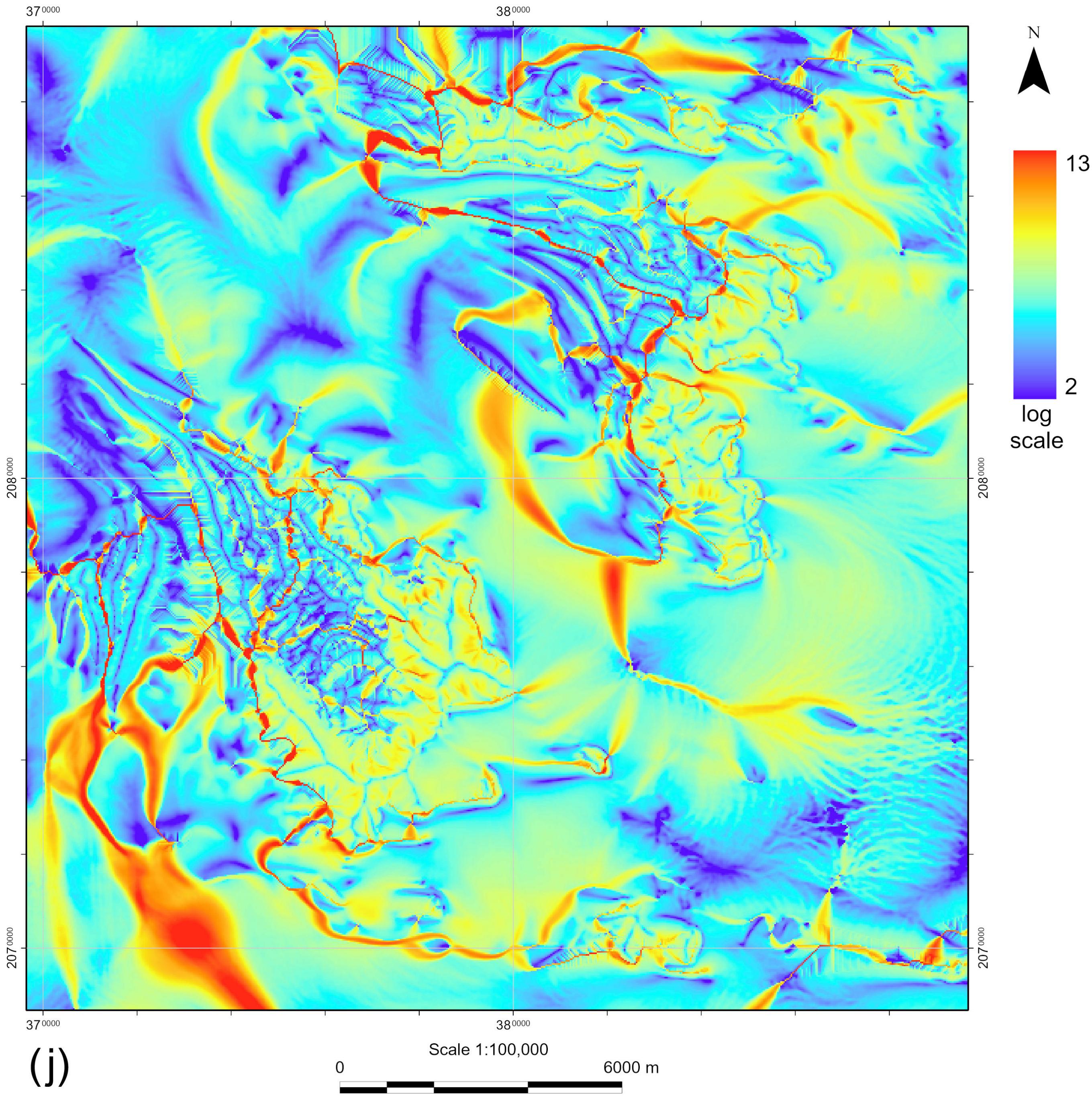

**Fig. 5, cont'd** Yamato (Queen Fabiola) Mountains, sheet I: ( j) Stream power index.

*(Continued)*

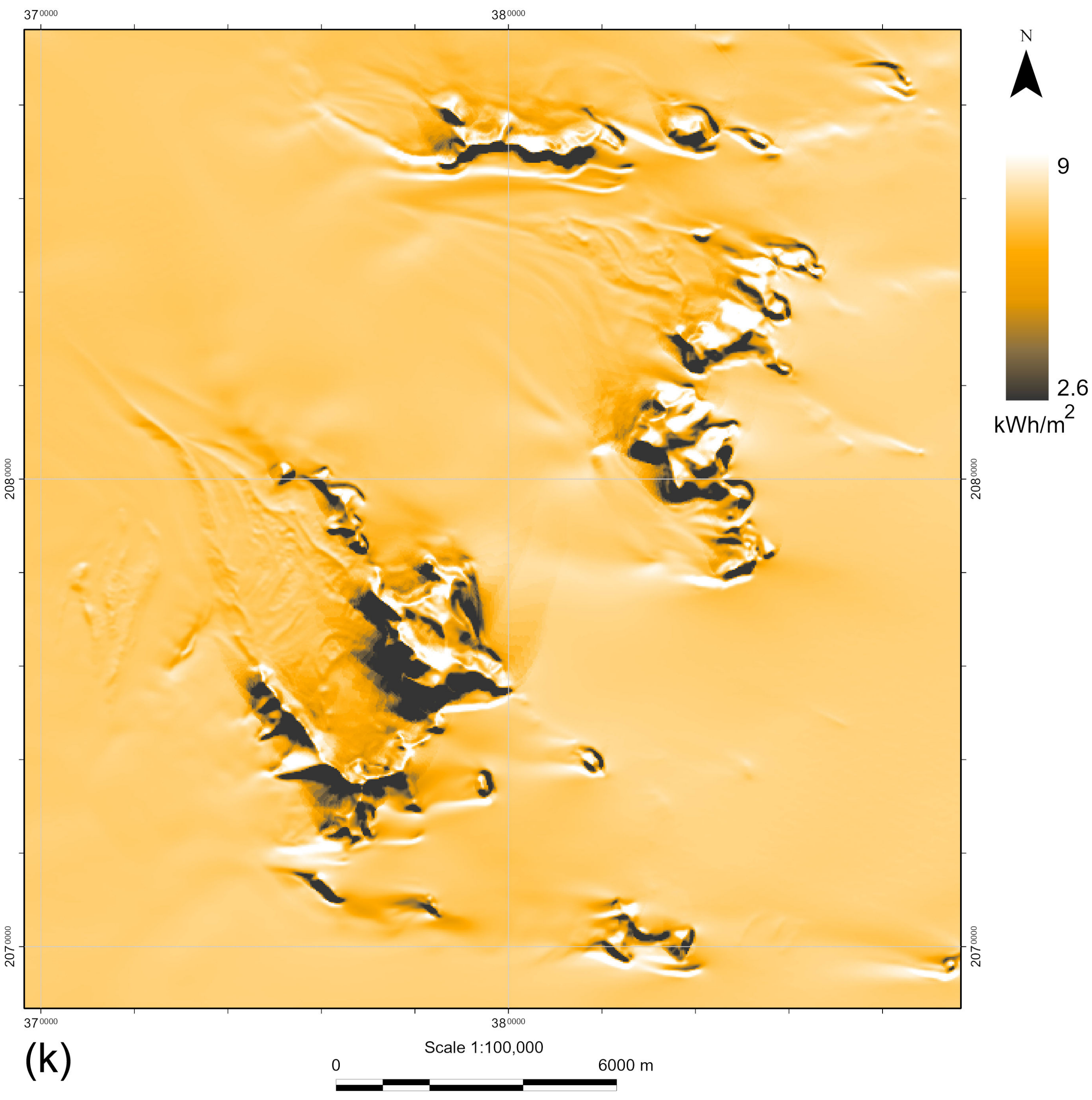

**Fig. 5, cont'd** Yamato (Queen Fabiola) Mountains, sheet I: (k) Total insolation.

*(Continued)*

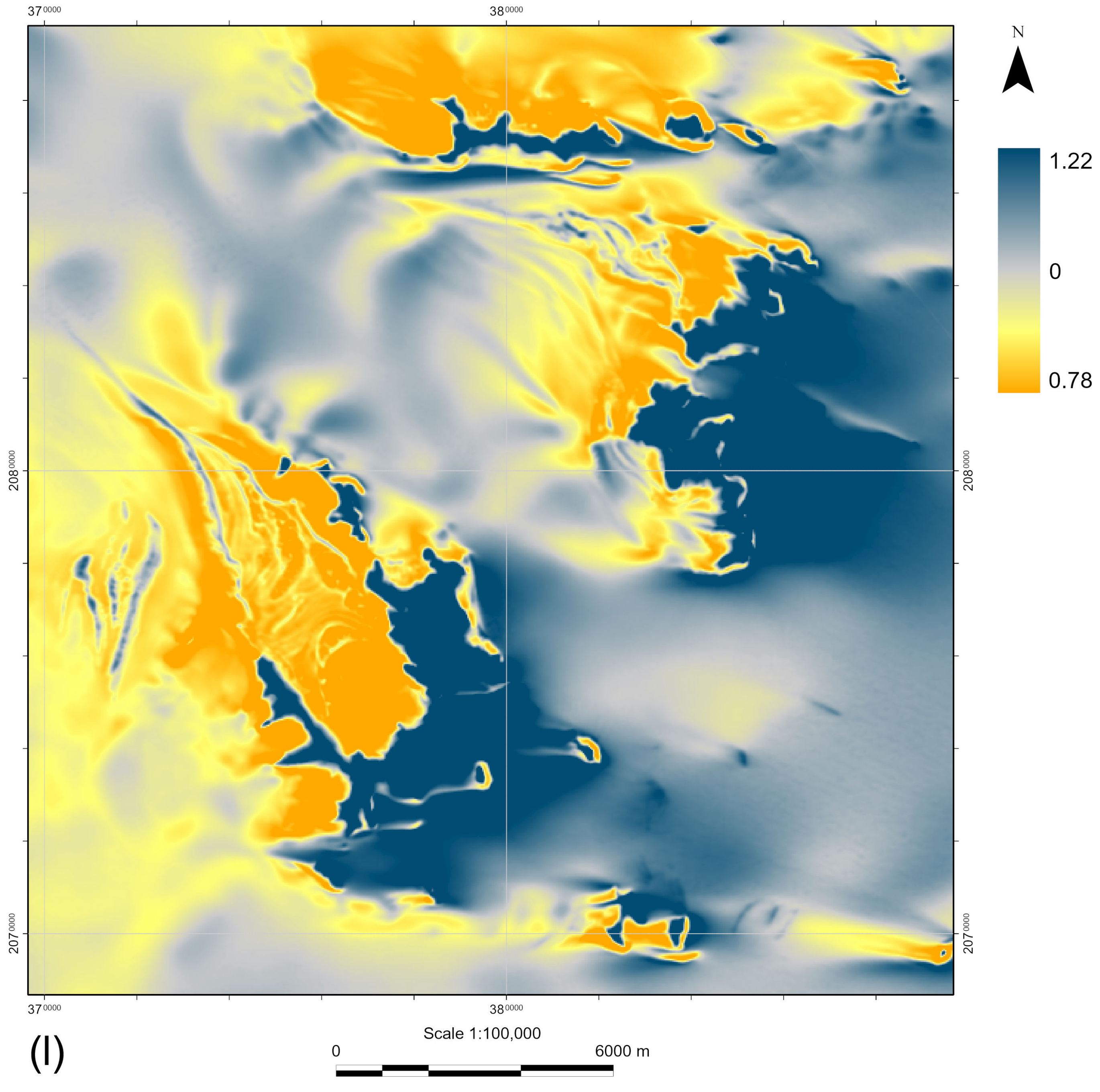

**Fig. 5, cont'd** Yamato (Queen Fabiola) Mountains, sheet I: (l) Wind exposition index.

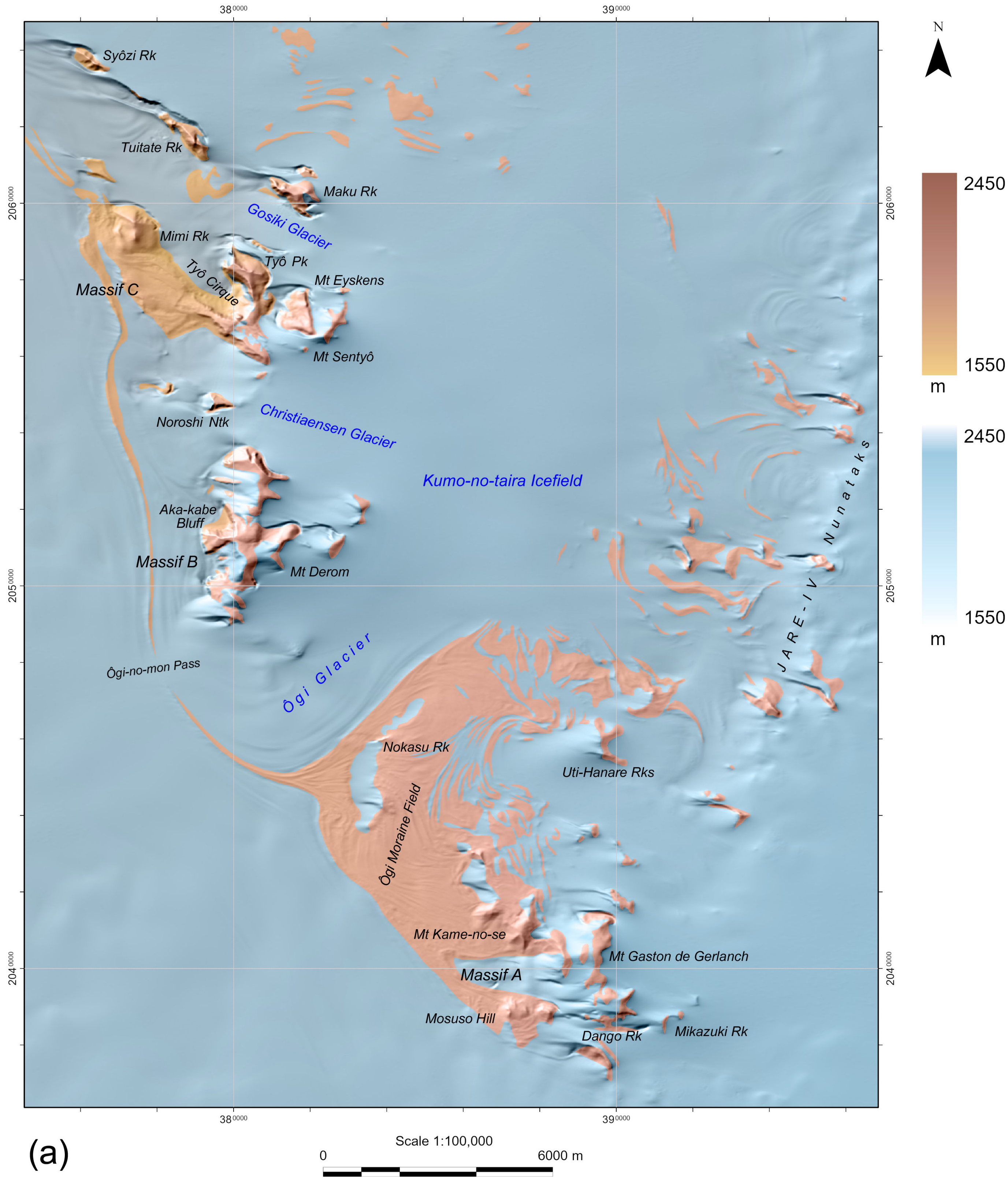

**Fig. 6** Yamato (Queen Fabiola) Mountains, sheet II: (a) Elevation.

*(Continued)*

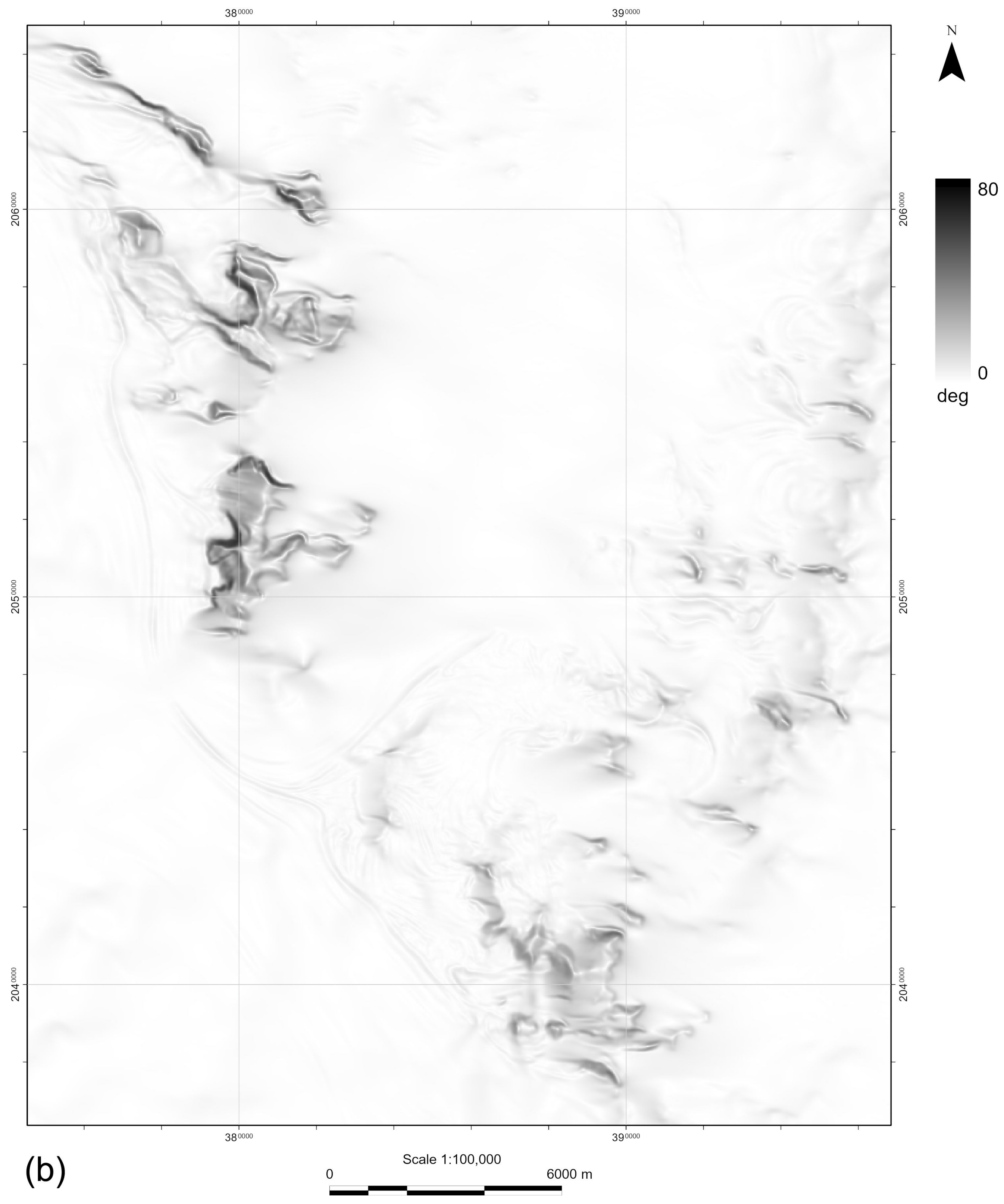

**Fig. 6, cont'd** Yamato (Queen Fabiola) Mountains, sheet II: (b) Slope.

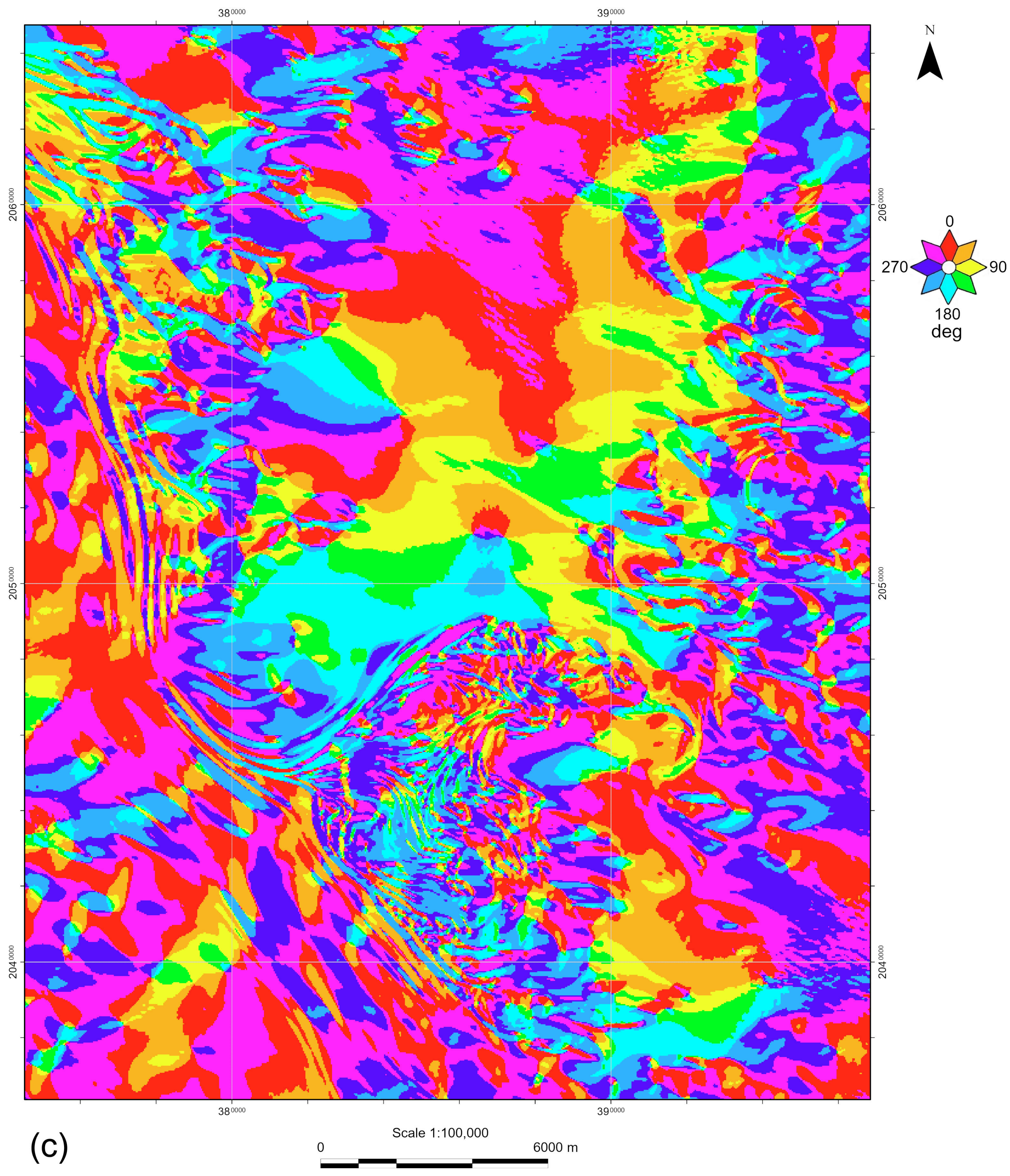

**Fig. 6, cont'd** Yamato (Queen Fabiola) Mountains, sheet II: (c) Aspect.

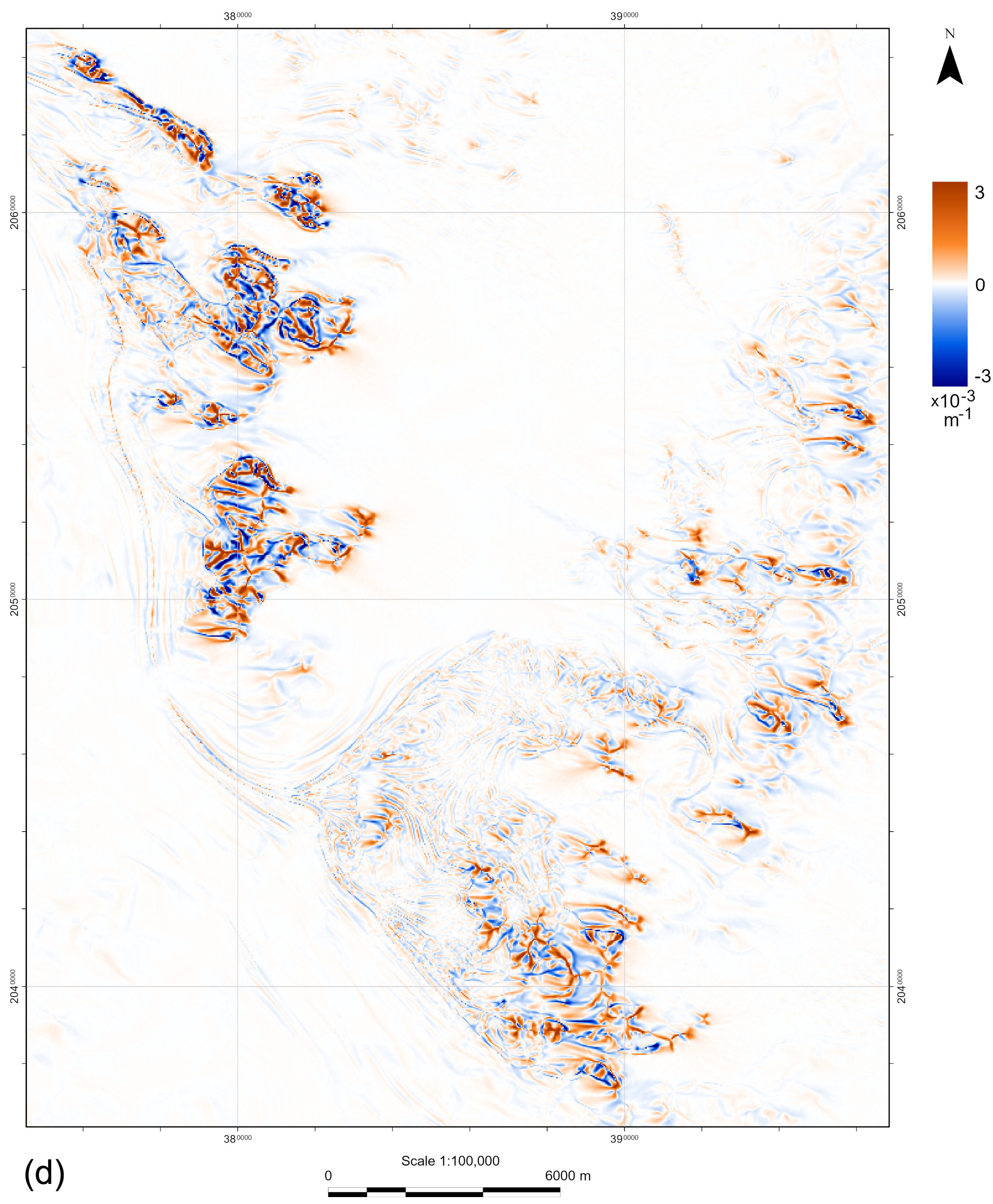

**Fig. 6, cont'd** Yamato (Queen Fabiola) Mountains, sheet II: (d) Horizontal curvature.

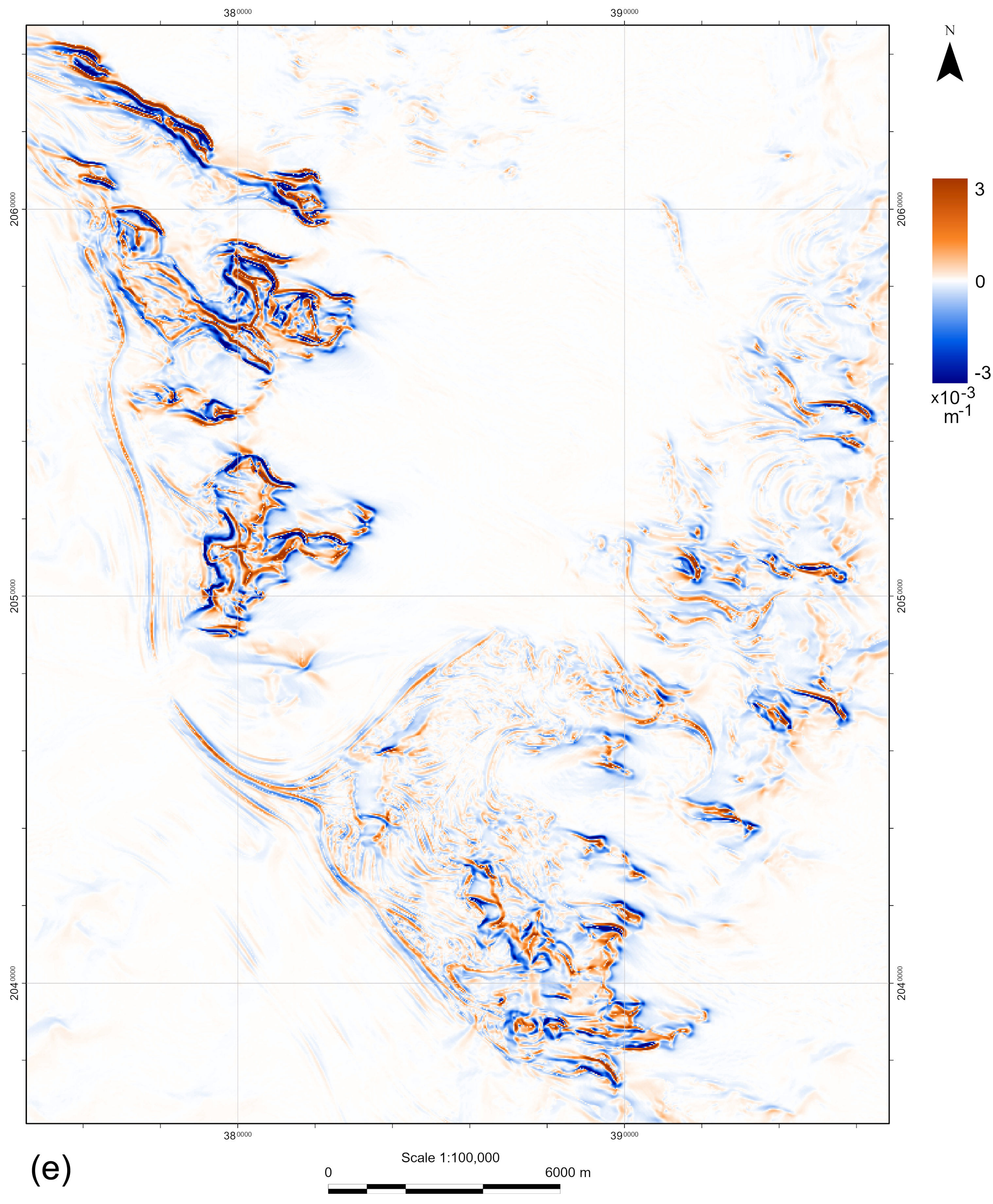

**Fig. 6, cont'd** Yamato (Queen Fabiola) Mountains, sheet II: (e) Vertical curvature.

*(Continued)*

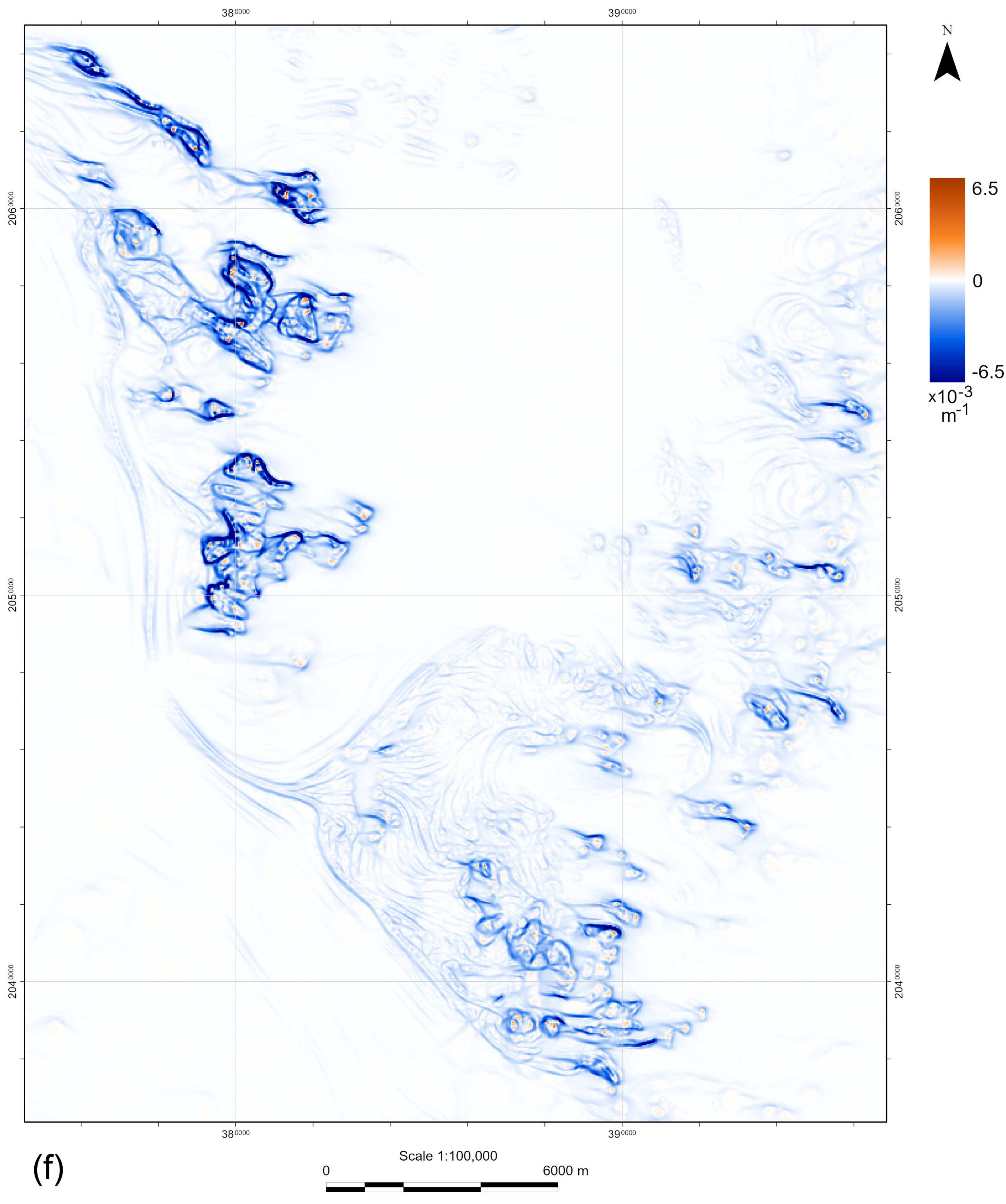

**Fig. 6, cont'd** Yamato (Queen Fabiola) Mountains, sheet II: (f) Minimal curvature.

*(Continued)*

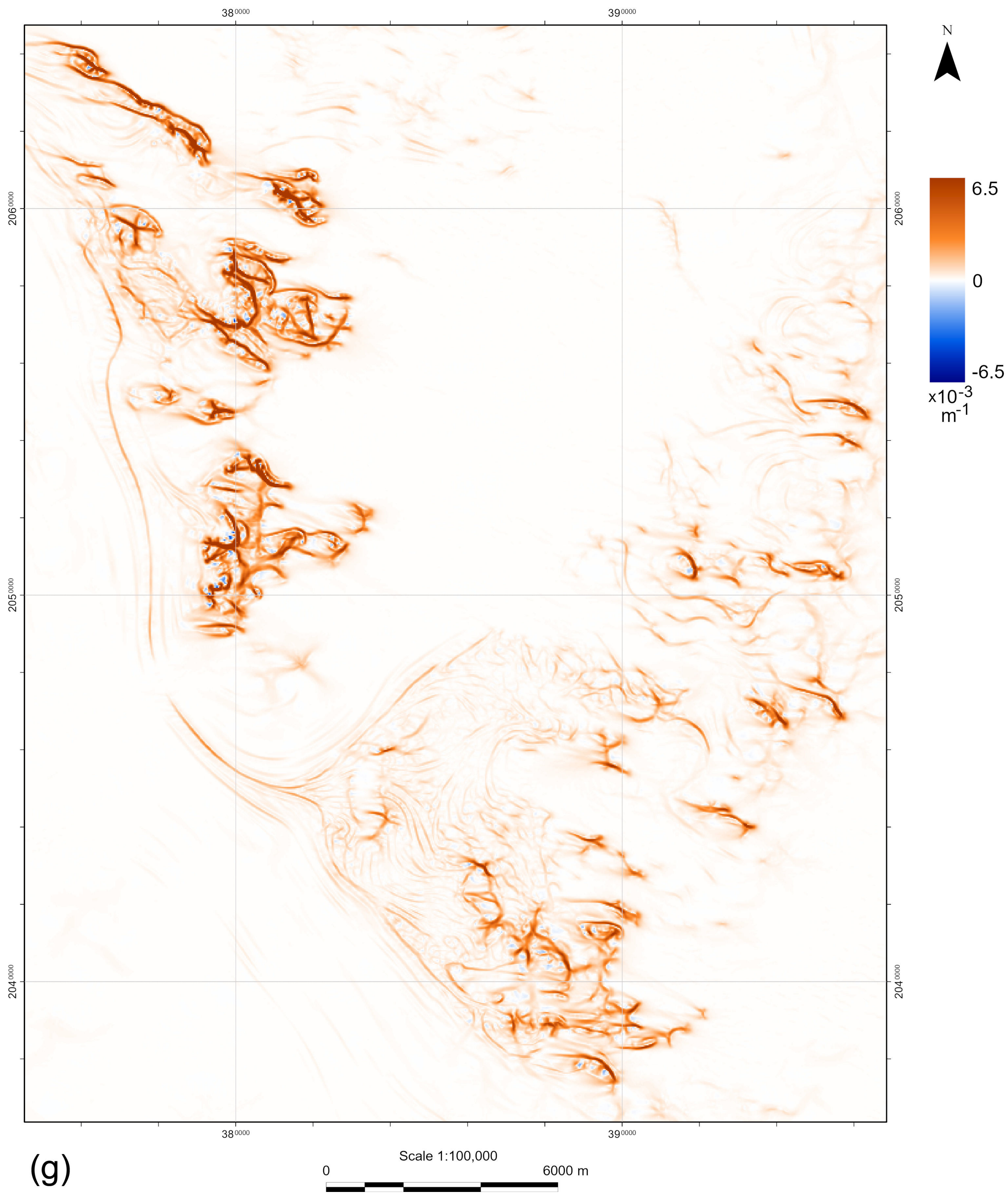

**Fig. 6, cont'd** Yamato (Queen Fabiola) Mountains, sheet II: (g) Maximal curvature.

*(Continued)*

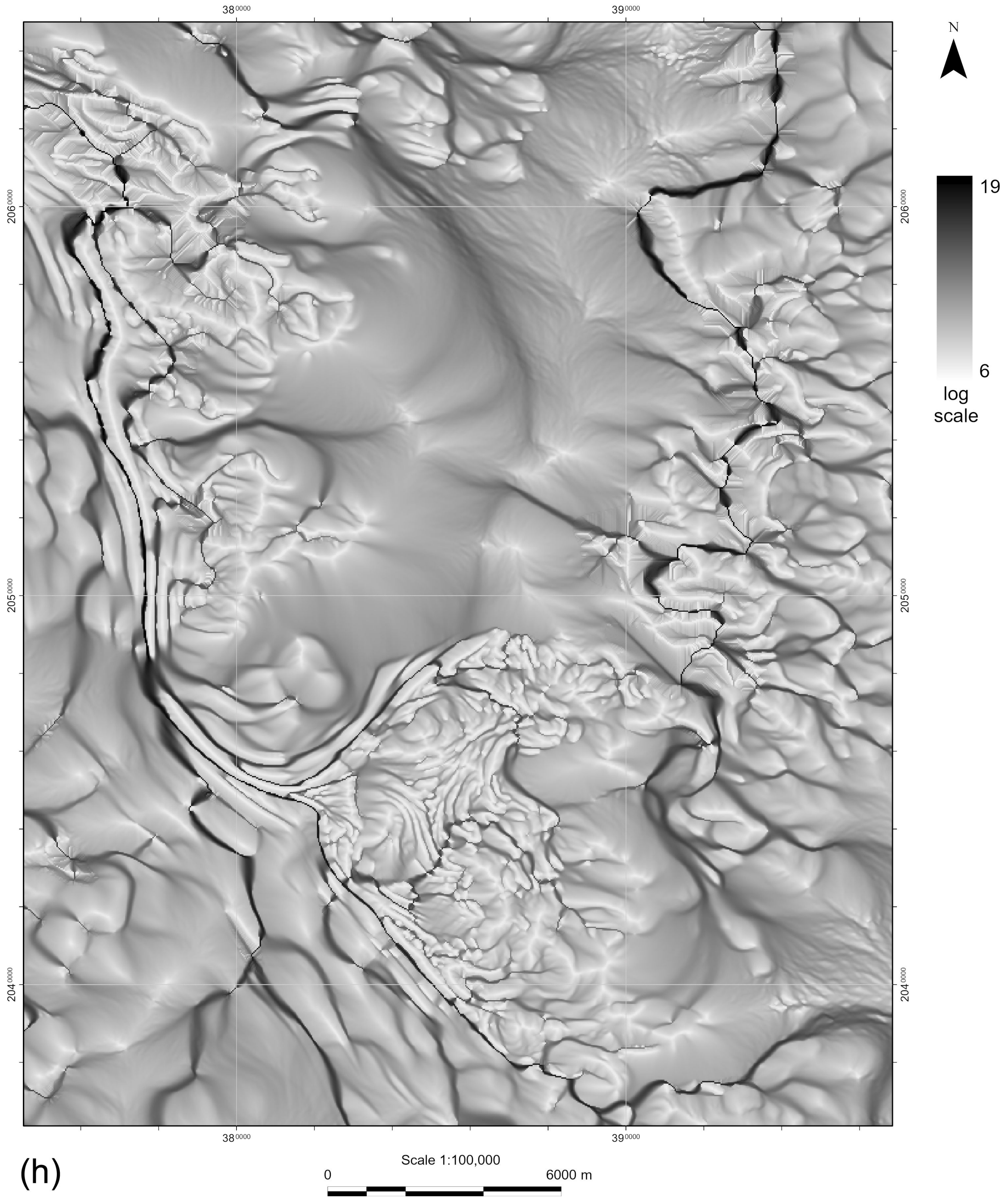

**Fig. 6, cont'd** Yamato (Queen Fabiola) Mountains, sheet II: (h) Catchment area.

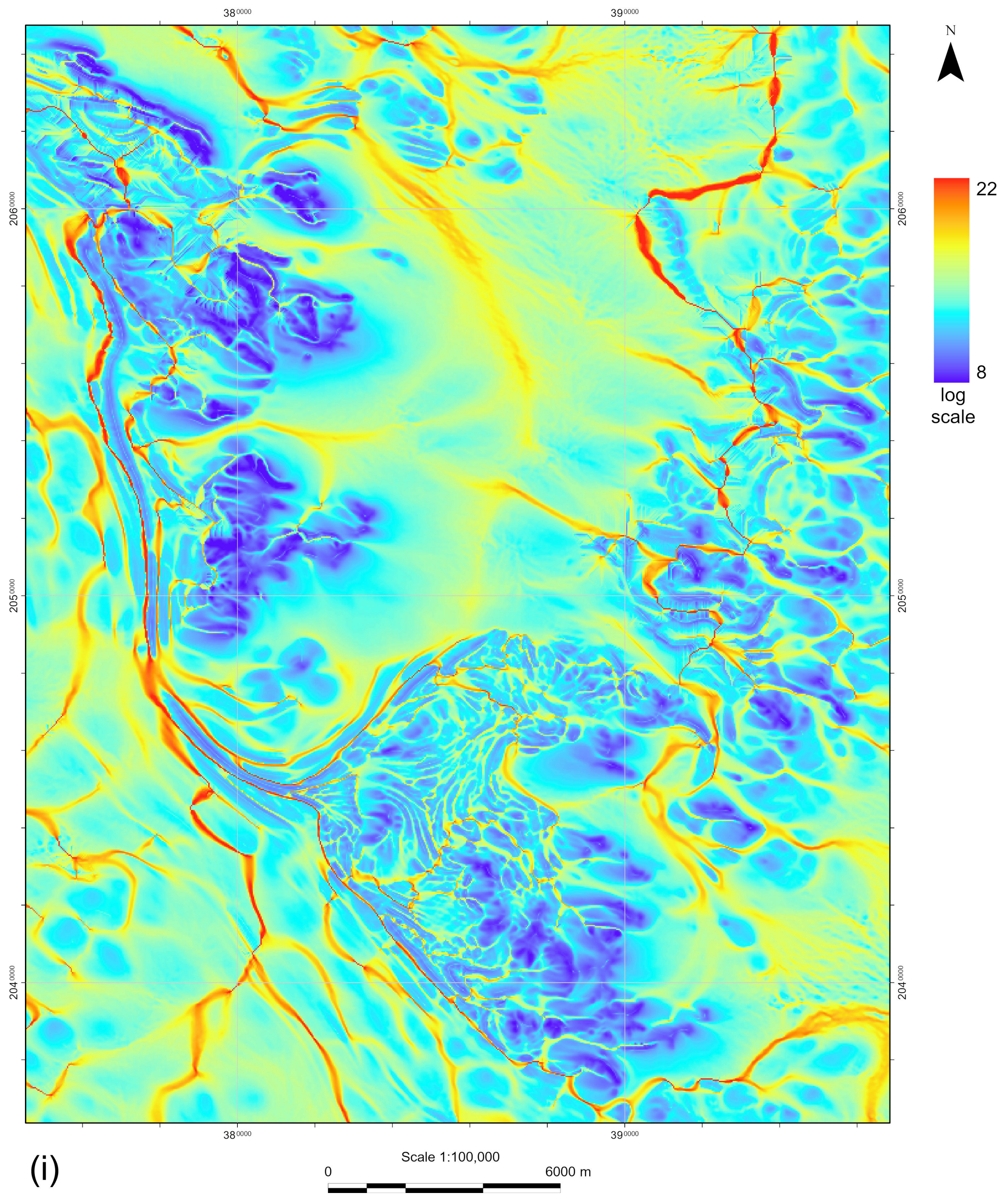

**Fig. 6, cont'd** Yamato (Queen Fabiola) Mountains, sheet II: (i) Topographic wetness index.

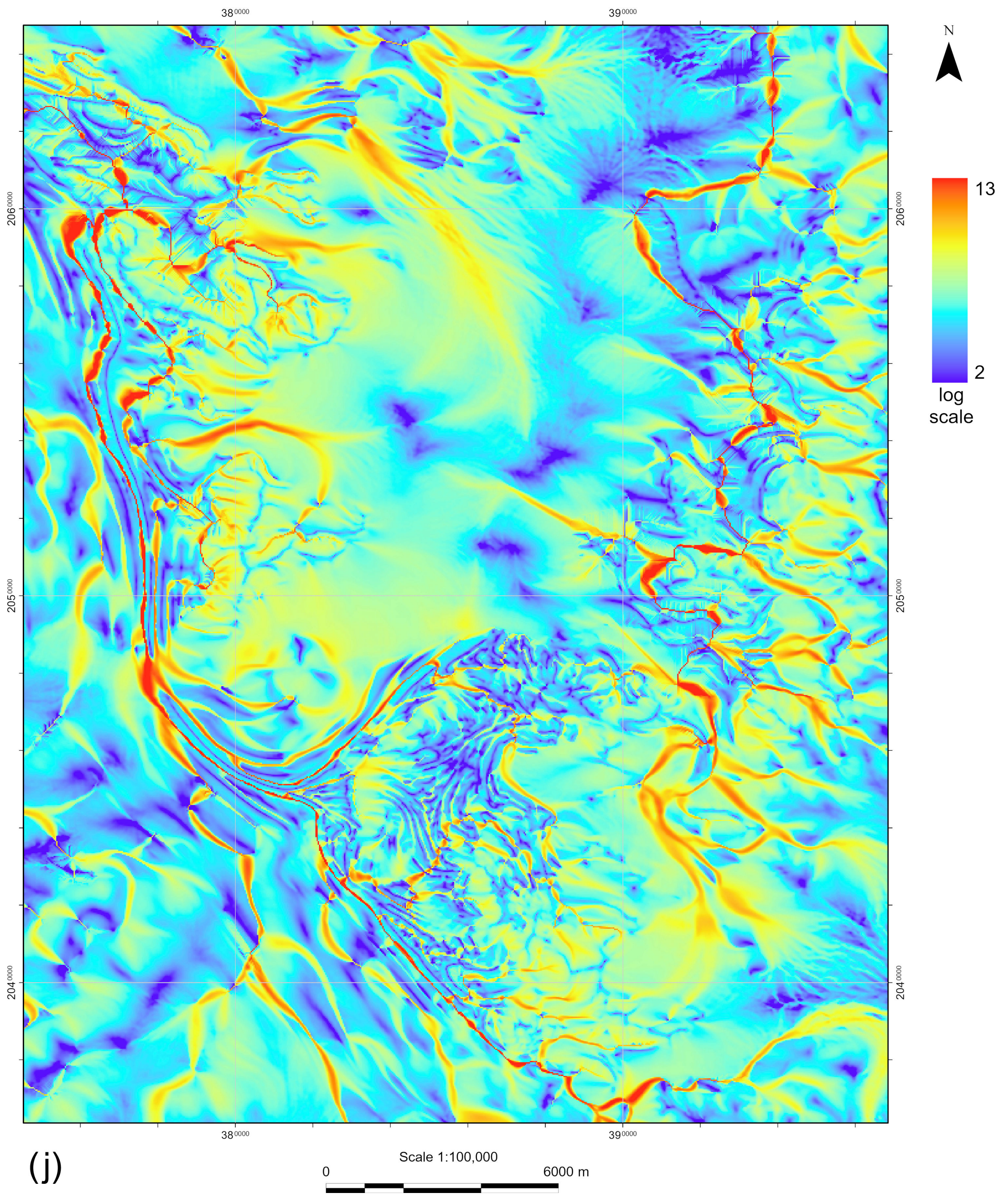

**Fig. 6, cont'd** Yamato (Queen Fabiola) Mountains, sheet II: ( j) Stream power index.

*(Continued)*

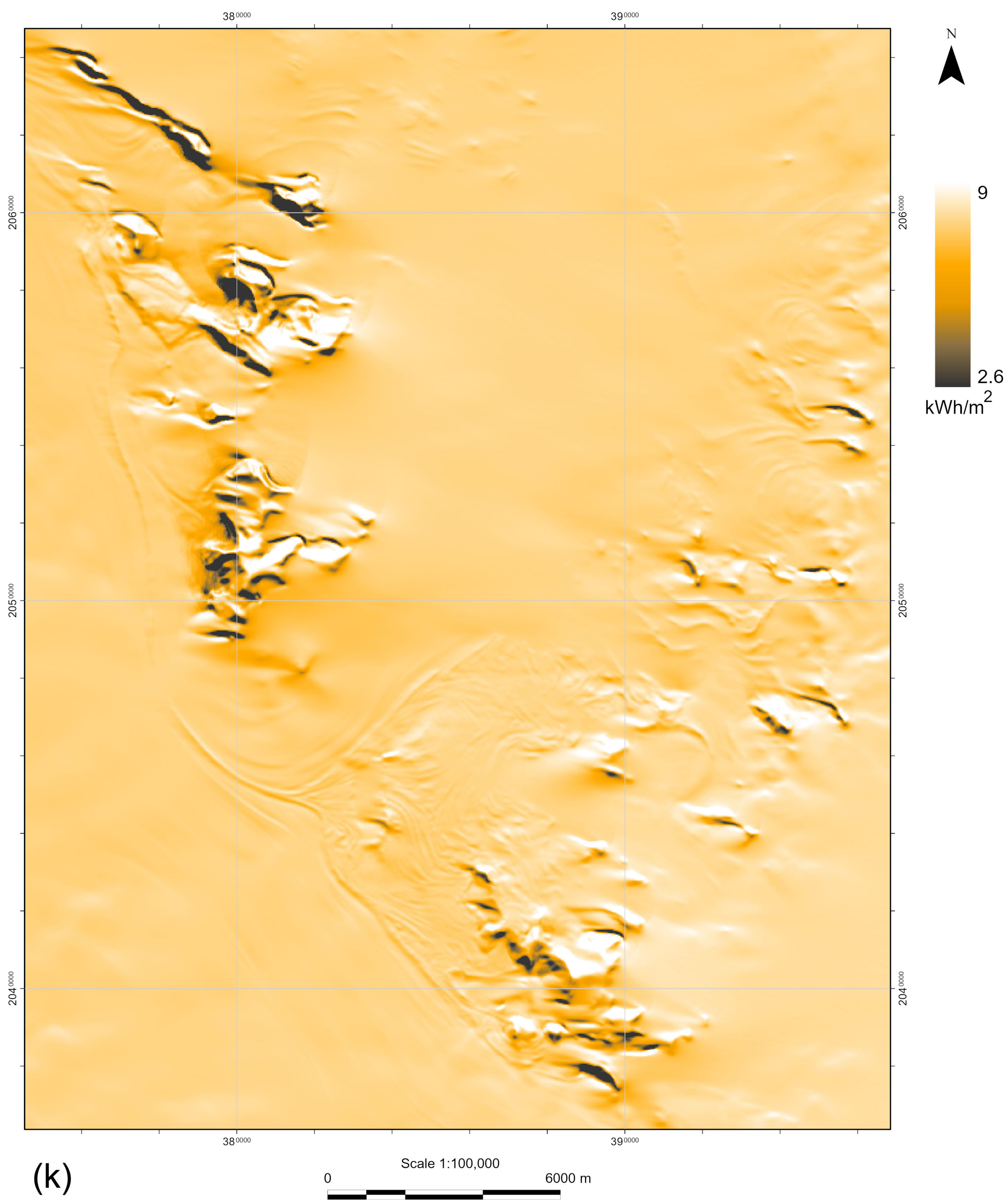

**Fig. 6, cont'd** Yamato (Queen Fabiola) Mountains, sheet II: (k) Total insolation.

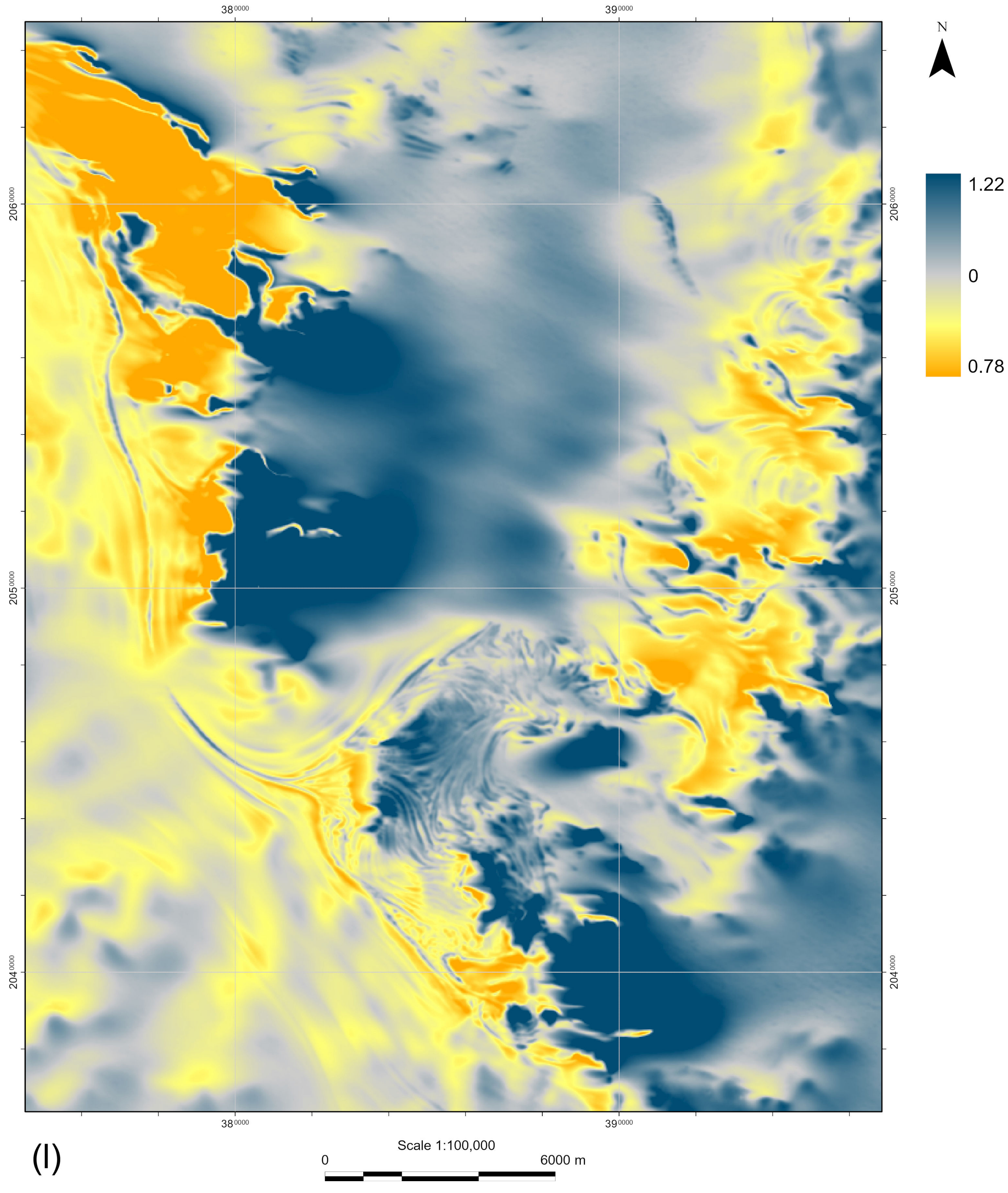

**Fig. 6, cont'd** Yamato (Queen Fabiola) Mountains, sheet II: (l) Wind exposition index.

## 5 Discussion

It is known[1] that topography controls the thermal, wind, and hydrological regimes of slopes, influencing therefore the distribution and properties of soils and vegetation (Huggett and Cheesman, 2002; Florinsky, 2025a).

The thermal regime of slopes depends in part on the incidence of solar rays to the land surface, so it depends on both *G* and *A*. *TIns* directly considers this incidence and better describes the thermal regime (Böhner and Antonić, 2009). Information on the differentiation of slopes by insolation level is critical to predict the spatial distribution of primitive soils and lower plants in ice-free areas. To refine such a prediction, one can use *TWI* digital models describing topographic prerequisites of water migration and accumulation.

A further refinement of such a prediction can be done with *WEx* data. *WEx* digital models are utilized to identify areas affected by and protected from the wind impact (Böhner and Antonić, 2009; Florinsky, 2025a, chap. 2). In periglacial Antarctic landscapes, one of the main meteorological factors determining the microclimate is katabatic wind. Thus, *WEx* maps may be of great importance for modeling the distribution of primitive soil and lower plants, predicting the differentiation of snow accumulation in ice-free areas, and determining the optimal location of buildings and infrastructure at polar stations.

$k_h$ maps display the distribution of convergence and divergence areas of surface flows. Geomorphologically, these are spurs of valleys and crests, respectively. The combination of convergence and divergence areas creates an image of the flow structures (Florinsky, 2025a, chap. 2). $k_v$ maps show the distribution of relative deceleration and acceleration areas of surface flows. Geomorphologically, these maps represent cliffs, scarps, terrace edges, and other similar landforms or their elements with sharp bends in the slope profile (Florinsky, 2025a, chap. 2). In this regard, $k_h$ and $k_v$ digital models may be useful in geomorphological and hydrological studies of the ice-free areas.

Combination of $k_h$ and $k_v$ digital models allows revealing relative accumulation zones of surface flows (Florinsky, 2025a, chap. 2). These zones, marked by both $k_h < 0$ and $k_v < 0$, coincide with the fault intersection sites and are characterized by increased rock fragmentation and permeability. Within these zones, one can observe an interaction and exchange between two types of substance flows: (a) lateral, gravity-driven substance flows moved along the land surface and in the near-surface layer, such as water, dissolved and suspended substances, and (b) vertical, upward substance flows, such as fluids, groundwater of different mineralization and temperature (Florinsky, 2025a, chap. 15). Maps of accumulation zones may be useful for geochemical studies in the ice-free areas.

$k_{max}$ and $k_{min}$ maps are informative in terms of structural geology because they clearly display elongated linear landforms (Florinsky, 2017, 2025a). In Antarctica, such lineaments can be interpreted as a reflection of the local fault and fracture network, which topographic manifestation has been amplified by erosional, exaration, and nival processes (Florinsky, 2023b; Florinsky and Zharnova, 2025a). Thus, researchers are able to use $k_{max}$ and $k_{min}$ models for compiling lineament maps and comparing them with other geological data.

*CA* maps can be used to identify the fine flow structure of drainage basins and then to incorporate this information into geochemical and hydrological analysis. *TWI* digital models can be applied for prediction of the ground moisture content in the ice-free areas as well as for forecasting the spatial distribution of snow puddles on adjacent glaciers in summer. *SPI* data can be useful for prediction of slope erosion in the ice-free areas as well as erosion of snow cover and ice by meltwater flows on adjacent glaciers in summer.

[1] This one-page explanation of the usefulness and possible application of morphometric models and maps in Antarctic research has earlier been published in our papers on ice-free geomorphometry of Antarctica (Florinsky, 2025b; Florinsky and Zharnova, 2025b, 2025c, 2025d, 2025e).

## 6 Conclusions

We performed geomorphometric modeling of the two, partly ice-free mountainous areas of the eastern Queen Maud Land, East Antarctica. For the first time, we created series of morphometric models and maps for the Belgica Mountains and the Yamato (Queen Fabiola) Mountains.

The obtained maps rigorously, quantitatively, and reproducibly describe the ice-free topography of these areas. New morphometric data can be used in further geological, geomorphological, glaciological, hydrological, and ecological studies of these Antarctic mountains.

The study was performed within the framework of the project for creating a physical geographical thematic scientific reference geomorphometric atlas of ice-free areas of Antarctica (Florinsky, 2024, 2025b).